\documentclass[fleqn,usenatbib]{mnras}
\usepackage{graphicx}
\usepackage{amssymb}
\usepackage{amsmath}
\usepackage{mathptmx}
\usepackage{array,booktabs}
\usepackage[normalem]{ulem}

\newcommand{\approptoinn}[2]{\mathrel{\vcenter{
  \offinterlineskip\halign{\hfil$##$\cr
    #1\propto\cr\noalign{\kern2pt}#1\sim\cr\noalign{\kern-2pt}}}}}

\usepackage{multirow}

\newcolumntype{P}[1]{>{\centering\arraybackslash}p{#1}}
\newcolumntype{M}[1]{>{\centering\arraybackslash}m{#1}}

\graphicspath{{./}{figures/}}

\newcommand{\Rt}{\,{\tilde{R}}}
\newcommand{\Rs}{\,{R_{\rm sp}}}

\newcommand{\Gba}{\,{\Gamma\beta_{\rm a}}}
\newcommand{\Ga}{\,{\Gamma_{\rm a}}}
\newcommand{\Gs}{\,{\Gamma_{\rm sp}}}
\newcommand{\Gbs}{\,{\Gamma\beta_{\rm sp}}}
\newcommand{\Gb}{\,{\Gamma\beta}}
\newcommand{\qNR}{\,{\theta_{\rm NR}}}
\newcommand{\qco}{\,{\theta_{\rm c,0}}}
\newcommand{\qc}{\,{\theta_{\rm c}}}
\newcommand{\qsp}{\,{\theta_{\rm c,sp}}}
\newcommand{\Eiso}{\left<E_{iso}\right>}

\title[Evolution of Relativistic Jetted Blast Waves]{The Structure and Evolution of Relativistic Jetted Blast Waves}
\author[Govreen-Segal \& Nakar]{
    Taya Govreen-Segal\thanks{taya@govreensegal.com}, Ehud Nakar
	\\
	{School of Physics and Astronomy, Tel Aviv University, Tel Aviv 6997801, Israel}
}
\pubyear{2023}
\begin{document}
	\label{firstpage}
	\pagerange{\pageref{firstpage}--\pageref{lastpage}}
	\maketitle

\begin{abstract}
We study, analytically and numerically, the structure and evolution of relativistic jetted blast waves that propagate in uniform media, such as those that generate afterglows of gamma-ray bursts. Similar to previous studies, we find that the evolution can be divided into two parts: (i) a pre-spreading phase, in which the jet core angle is roughly constant, $\theta_{c,0}$, and the shock Lorentz factor along the axis, $\Gamma_a$, evolves as a part of the Blandford-Mckee solution, and (ii) a spreading phase, in which $\Gamma_a$ drops exponentially with the radius and the core angle, $\theta_c$, grows rapidly. Nevertheless, the jet remains collimated during the relativistic phase, where $\theta_c(\Gamma_a\beta_a=1)\simeq 0.4\theta_{c,0}^{1/3}$. The transition between the phases takes place when $\Gamma_a\simeq 0.2\theta_{c,0}^{-1}$.
We find that the "wings" of jets with initial "narrow" structure ($\frac{d \log\,E_{iso}}{d\log\,\theta}<-3$ outside of the core, where $E_{iso}$ is isotropic equivalent energy), start evolving during the pre-spreading phase. By the spreading phase these jets evolve to a self-similar profile, which is independent of the initial structure, where in the wings $\Gamma(\theta)\propto\theta^{-1.5}$ and $E_{iso}(\theta)\propto \theta^{-2.6}$. Jets with initial "wide" structure roughly keep their initial profile during their entire evolution.
We provide analytic description of the jet lateral profile evolution for a range of initial structures, as well as the evolution of $\Gamma_a$ and $\theta_c$. 
For off-axis GRBs, we present a relation between the initial jet structure and the light curve rising phase.
Applying our model to GW170817, we find that initially the jet had $\theta_{c,0}=0.4-4.5~\deg$ and wings which are consistent with $E_{iso} \propto \theta^{-3}-\theta^{-4}$.
\end{abstract}

\section{Introduction} \label{sec:intro}
Afterglows of gamma-ray bursts (GRBs) of all types, as well as jetted tidal-disruption events, are well explained by the emission from a jetted blast wave propagating in the surrounding medium, emitting synchrotron radiation.  
While many aspects of the afterglow physics are well understood, the jet fluid dynamical evolution remains a challenge. This evolution has a significant impact on the observed emission for both on-, and off-axis observers. It determines the light-curve decline rates \citep[e.g.,][]{Sari1999,Granot2007}, the relation between the jet core angle measured from observations and it's initial opening angle, and may have a significant effect on the afterglow image \citep{Fernandez2022,GovreenSegal2023} and the evolution of light polarization \citep{Rossi2004}. For off-axis observers, the jet structure evolution also impacts the rising of the light-curve, and the timing and flux of the light-curve peak \citep[e.g.,][]{Ryan2020,Takahashi2021,Nakar2021}.

Given its importance, a significant effort has been made to understand the jet hydrodynamic evolution. This effort focused mainly on the characterization of two global quantities - the blast wave's characteristic Lorentz factor, $\Gamma$, and its opening angle $\theta_c$. The key property of relativistic jets in this regard is that sideways motion of fluid elements and sound waves, as measured in the observer frame, is at most at a velocity of about $1/\Gamma$ of the speed of light. This implies that the jet is causally connected over an angular scale of $\sim \Gamma^{-1}$. Therefore, as long as $\Gamma \gg 1/\theta_c$ every point on the core evolves as a part of a spherical blast wave, according to the \cite{Blandford1976} solution \citep{Piran1995}. However, once $\Gamma \sim 1/\theta_c$ the evolution is expected to change as sideways spreading becomes important. The first to discuss this change was \cite{Rhoads1997,Rhoads1999}, which showed that if the core expands at the speed of sound then once $\Gamma \sim 1/\theta_c$ the evolution changes so $\Gamma$ drops exponentially with the radius while $\theta_c$ grows exponentially. Based on these considerations \cite{Sari1999} approximated that at any time after spreading starts  $\Gamma \theta_c \sim 1$, and used this approximation to predict the post jet-break light curve.

Early numerical studies of the evolution \citep{Granot2001,Meliani2007,Zhang2009,vanEerten2010,Wygoda2011,DeColle2012} proved stimulating jets to be extremely challenging due to the large dynamic range between the shock radius, $R$, and the width of the shocked matter $\sim R/\Gamma^2$. This scale separation limited the initial shock Lorentz factor in most of these simulations to be moderate (in the range of 20-60) and the corresponding initial opening angle to be 0.1-0.2 rad. As a result the dynamical range over which the relativistic spreading phase could have been simulated was small, making the interpretation of the results difficult. For example, \cite{Zhang2009,vanEerten2012} and \cite{Wygoda2011} studied the jet spreading numerically, and although their numerical results were similar, the two former studies concluded that the jet spreading is logarithmic while the latter argued for a short phase of exponential spreading followed by slower evolution. 

 In view of the numerical results, \cite{Granot2012} revisited the analytic solution for the evolution of $\Gamma$ and $\theta_c$ during the spreading phase. They used the assumption that the jet is a top hat at all times (i.e., it is uniform over its opening angle and there is no energy outside of that angle), and considered different approximations for assessing the amount of mass swept up by the shock as it propagates. For each approximation they used conservation of energy and momentum to calculate the evolution of $\Gamma$ and $\theta_c$. They also extended the solutions to the Newtonian regime. Their solution suggested (for all approximations) that the spreading is slower than the naive prediction of \cite{Rhoads1999}, similarly to what was seen in the simulations. Yet, in their solutions spreading is exponential during the ultra-relativistic phase. They concluded that unless the initial opening angle of the jet is smaller than about 0.05 rad, the dynamic range is too small to properly identify the relativistic exponential spreading phase.

Recently,  \cite{Duffell2018} carried out numerical simulations with a larger dynamic range. They simulated jets with several initial opening angles, the smallest being about $0.04$ rad and the highest initial Lorentz factor being almost 100. They calibrated one of the approximations suggested by \cite{Granot2012}, which includes the transition to the mildly relativistic phase, to obtain an analytic model that fits the evolution of $\Gamma$ and $\theta_c$ in their simulations. They find that while the jet spreading is slower than the \cite{Rhoads1999} approximation, it does seem to spread rapidly during the short relativistic phase in their simulations.

To conclude, while it is clear that the jet spreads during the relativistic phase, it is unclear to what extent. There is a semi-analytic description that fits a limited number of simulations, but there is no simple analytic description for the evolution of $\Gamma$ and $\qc$ during the relativistic phase, or for its dependence on the jet initial conditions. Obtaining such a solution is one of the goals of this paper.  

Another important aspect of the jets' dynamics is the evolution of its angular structure. It is clear that realistic jets have some initial angular structure when they emerge from the stellar envelope in the case of long GRBs or the merger ejecta in the case of short GRBs. Moreover, regions outside of the core, where the Lorentz factor is much lower, must evolve also before the onset of core spreading. The evolution of the jet structure outside the jet core is of special interest in the context of GRBs observed off-axis, as the structure outside the core dominates the rising phase of the light curve. Since the jet structure at a given time of its propagation is a result of the initial conditions as well as the jet evolution at earlier times, we need to understand the structure evolution of jets with various initial structures. So far the hydrodynamic evolution of the jet structure was not studied in detail.   \cite{Ramandeep2019} made an attempt to test whether a jet with initial condition of a top hat can evolve to a shallow enough structure so it may explain the rising phase of the afterglow of GW170817. They suggested that it may be possible,  although their numerical resolution was not sufficient to verify this claim. This exemplifies the type of questions that we would like to better understand.

In this work we study the hydrodynamic evolution of relativistic jetted blast wave that propagates in a constant density medium. We examine global properties, such as $\Gamma$ and $\theta_c$ as well as structure evolution. Our goal is to understand the dependence of the jet evolution on the jet's initial structure and to provide a simple analytic description of this evolution. For that, we use analytic considerations as well as numerical simulations with the public code GAMMA \citep{Ayache2022}. This code enables us to start our simulations with a matter Lorentz factor of 100-400 (the shock Lorentz factor is larger by a factor of $\sqrt{2}$) and an initial opening angle as small as 0.005 rad, so there is ample separation between the beginning of the spreading phase and the Newtonian phase. 

We proceed as follows: in \S\ref{sec:analytical} we give an analytic model to the evolution of $\Gamma$ and $\theta_c$. In \ref{sec:simulations}, we present our numerical simulations. We test and calibrate the analytic model of $\Gamma$ and $\theta_c$ and study, numerically, the evolution of the jet lateral structure. In \S\ref{sec:emission} we discuss the implications of our finding to the afterglow emission.  This  includes a comparison of light curves from the full hydrodynamic simulations to semi-analytic light curve models and an analytic formula that relates the slope of the rising phase to the initial jet structure. In this context we give in appendix \ref{appendix:structure_approx} an analytic approximation of the structure evolution of power-law jets, which can be used to obtain improved modeling of the light curves from such jets. In \S\ref{sec:170817} we use our results to derive constraints on the initial opening angle and lateral structure of the jet in GW170817. We summarize our results in  \S\ref{sec:Conclusions}.

\section{Analytic model for $\Gamma$ and $\theta_c$}\label{sec:analytical}
We start by describing our analytical expectations for the fluid dynamical evolution of a jet, focusing on the evolution of the typical angular scale, and the typical Lorentz factor. 
Consider a jet, with a core angle $\theta_c$ (to be defined rigorously later), and a typical shock Lorentz factor $\Gamma_a$, defined as the Lorentz factor along the jet axis (here and in the rest of the paper, we use the subscript "a" to denote the value along the jet axis). The fluid behind the shock is in causal contact on an angular scale of $\sim\frac{1}{\Gamma}$. Therefore, while $\Gamma_a\theta_c\gg1$, different regions in the core are not yet causally connected, and the jet core evolves as part of a Blandford-Mckee (BM) solution \citep{Blandford1976} with the local initial isotropic equivalent energy, which we denote $E_{iso,i}$. Hence, the Lorentz factor of the shock along the jet axis can be expressed as:
\begin{equation}\label{eq:Gbm}
    \Gamma_a = \left(\frac{17}{8\pi}\frac{E_{iso,a,i}}{\rho c^2 R_a^3}\right)^{1/2}~~~ ; ~~~ \Gamma_a\theta_c\gg1,
\end{equation}
where $\rho$ is the external density, $R_a$ is the blast wave radius along the axis, $c$ is the speed of light and $E_{iso,a,i}$ is the initial isotropic equivalent energy along the axis. During this phase, we expect the jet core angle to be roughly constant.

Once the edges of the core become causally connected, when $\Gamma_a\theta_c\simeq1$, the core starts to spread. \cite{Keshet2015} (see also \citealt{Gruzinov2007}) showed that the axisymmetric fluid equations obey a scaling that is indicative of the existence of a self-similar solution. They showed that in such a solution, scaling laws indicate that the Lorentz factor in the jet core decreases more rapidly than a power law and they assumed that it decreases exponentially. In the numerical section, we indeed find an evolution that is consistent with an attractive self-similar solution in which the Lorentz factor decreases exponentially\footnote{While we choose the same scaling as \cite{Keshet2015} for $\Gamma_a$, the scaling we adopt for $\theta_c$ and for $ct-R_a$ is different than the one they chose. Note, however, that our choice is also consistent with their scaling argument.}, although naturally, we cannot prove numerically that the solution is indeed fully self-similar. 

We choose to formulate the spreading phase with an exponential scaling in radius rather than in time, as we find that in that manner it remains relevant also in the mildly relativistic case. For this reason, we also write the expression in terms of the proper velocity. Denoting $\Gbs$ and $\Rs$ as the values of $\Gba$ and $R_a$ (the shock proper velocity and the shock radius along the axis, respectively) at the transition between the BM and spreading regimes, the  proper velocity in the spreading regime is given by:
\begin{equation}\label{eq:Gsp}
    \Gba=\Gbs\exp\left(-\frac{R_a-\Rs}{\Rt}\right)~~~;~~~\Gba\theta\ll1
\end{equation}
where $\Rt$ is the length scale over which $\Gba$ decreases by an e-fold during the spreading phase. To find $\Rt$ we note that it should be of the same order of magnitude as $\Rs$, where the relations of the BM solutions are applicable and at the same time $\Gamma_a \sim \theta_c^{-1}$. Thus, $E_{iso,a} \sim \Rt^3 \Gamma_a^2 \rho c^2  \sim \Rt^3 \theta_c^{-2} \rho c^2$, and since the total blast wave energy satisfies $E_{tot} \sim E_{iso,a}\theta_c^2$ we obtain $\Rt \sim (E_{tot}/\rho c^2)^{1/3}$, and introducing a calibration constant $\tilde{C}$:
\begin{equation}\label{eq:Rt}
    \Rt =  \tilde{C} \left(\frac{E_{tot}}{\rho c^2}\right)^{1/3}.
\end{equation}

We would now like to determine the evolution of $\theta_c$ during the spreading phase.
The relation between $\Gba$ and $\qc$ generally depends on the mass and energy transfer within the complete jet structure, which is difficult to model analytically. Under the assumption that the solution is self-similar, the angular structure of various properties describing the shock depends only on $\frac{\theta}{\theta_c}$, and therefore, the core energy is a constant fraction of the jet energy.
\begin{equation}\label{eq:Ec}
    E_c \propto E_{tot} = C_{tot} \rho c^2 R_a^3\theta_c^2\Gba^2~~~;~~~R_a > R_{sp}.
\end{equation}
where 
$C_{tot}$ is a constant that depends on the exact self-similar structure\footnote{
In Eq. \ref{eq:Ec}, we assume that the shocked mass scales as $R^3$, rather than the lower powers of $R$ suggested in other works (e.g. one of the models by \citealt{Granot2012}). The reason is that we expect the blast wave to collect mass at all radii also outside of the core, by the wings, and then as the core spreads it transfers energy into this mass. By this method, the jet cannot "bypass" a significant part of the external matter in its wake. Note that while the scaling of the shocked mass in this equation is the same as for a non-spreading jet approximation, we do not expect the mass at every angle to be the mass collected by that angle, as it is likely that within the jet structure mass flows from the core to the wings. }.
We use Eq. \eqref{eq:Ec} to determine $\theta_c$ by means of energy conservation:
\begin{equation}\label{eq:qc}
    \theta_c \approx \qsp\left(\frac{R_{a}}{R_{sp}}\right)^{-\frac{3}{2}}e^{\frac{R_{a}-R_{sp}}{\tilde{R}}}~~~;~~~ R_a > \Rs
\end{equation}
where $\qsp = \qc(R_a=\Rs)$. 
We can now determine $\qsp$, by solving $\Gbs$ from Eq. \eqref{eq:Ec}, equating it with Eq. \eqref{eq:Gbm} and assuming $\Gs\gg1$, i.e; $\Gbs\simeq\Gs$, we find that:
\begin{equation}\label{eq:qsp}
    \qsp = \left(\frac{8\pi}{17 C_{tot}}\frac{E_{tot}}{E_{iso,i}}\right)^{\frac{1}{2}}
\end{equation}
We can use Eq. \eqref{eq:Ec} to find $\Rs$:
\begin{equation}\label{eq:Rsp}
    \Rs = C_{sp}^{-\frac{2}{3}}\left(\frac{1}{C_{tot}}\frac{E_{tot}}{\rho c^{2}}\right)^{\frac{1}{3}} = \tilde{C}^{-1}C_{sp}^{-\frac{2}{3}} C_{tot}^{-\frac{1} {3}} \Rt,
\end{equation}
here $C_{sp}=\Gbs\qsp\sim1$, and its exact value will be calibrated numerically. 

We obtained an analytic model of $\Gamma_a$ and $\qc$ for which we need to find three order unity calibration constants: $\tilde{C}$ - the normalization of the e-folding radius, $C_{tot}$ - the total energy normalization constant, and $C_{sp}=\Gbs\qsp$. In the following section, we validate the analytic model and find these constants 
 (for our specific definition of $\theta_c$) using numerical simulations. Their value can depended on the initial jet structure, where we find that for top hat jets $\tilde{C}=0.65$, $C_{tot}=0.54$, and $C_{sp}=0.38$. For these parameters $\Rs = 3.6 \Rt$. To apply the model to structured jets, additional generalizations are needed, these are discussed in \S\ref{sec:structured}. For convenience, we provide a glossary in table \ref{tab:glossary}.

 \begin{table*}
     \centering
\begin{tabular}{|c|l|}
\hline 
Symbol & Definition\tabularnewline
\hline 
\hline 
$\Gamma_{a}/\Gamma\beta_{a}$ & The shock Lorentz factor/proper velocity along the jet axis\tabularnewline
\hline 
$\Gamma\beta_{sp}$ & The shock proper velocity along the jet axis at the onset of spreading, $\Gba (\Rs)$\tabularnewline
\hline 
$\gamma_i$ & The initial Lorentz factor of the matter just behind the shock \tabularnewline
\hline 
$R_{a}$ & The shock radius along the jet axis\tabularnewline
\hline 
$R_{sp}$ & The shock radius at the transition from BM to exponential decay of $\Gba$, which we denote as the onset of the spreading phase\tabularnewline
\hline 
$\tilde{R}$ & The e-folding radius of the Lorentz factor deceleration during the spreading phase. In top hat jets $\Rs=3.6\Rt$. \tabularnewline
\hline 
$\theta_{c}$ & The jet core angle\tabularnewline
\hline 
$\theta_{i}$ & Opening angle set in the initial conditions, used to identify which
simulation is used.\tabularnewline
\hline 
$\theta_{c,0}$ & The core angle once the jet structure becomes stable and before the
onset of spreading\tabularnewline
\hline 
$\qsp$ & The core angle at the onset of the spreading phase, $\theta_c(\Rs)$. In top hat jets $\qsp \simeq 2\qco$.\tabularnewline
\hline 
$E_{iso,i}\left(\theta\right)$ & The initial isotropic equivalent energy\tabularnewline
\hline 
$\left\langle E_{iso}\right\rangle $ & The energy averaged isotropic equivalent energy\tabularnewline
\hline 
$E_{iso,i,a}$ & The initial isotropic equivalent energy along the axis\tabularnewline
\hline 
$E_{tot}$ & The total energy\tabularnewline
\hline 
$\rho$ & External density, assumed to be constant\tabularnewline
\hline 
\end{tabular}
     \caption{Glossary of our notations.}
     \label{tab:glossary}
 \end{table*}
 
\section{Numerical results}\label{sec:simulations}
\subsection{Simulation setup}
We use GAMMA\footnote{https://github.com/eliotayache/GAMMA} \citep{Ayache2022}, a code optimized for simulations of relativistic hydrodynamic blast waves, to run 2D jetted blast waves in spherical coordinates. We simulated jets in a uniform medium, setting the energy, external density, and initial radius so that the entire jet structure has passed its deceleration radius. As our focus is the ultra-relativistic evolution, we use an ideal gas equation of state with an adiabatic index $\hat{\gamma}=4/3$, though we test a few cases with an equation of state that transitions smoothly to $5/3$ in the sub-relativistic regime, finding it has minor effects on the evolution during the mildly relativistic phase (see discussion in Appendix \ref{appendix: EOS}). The initial jet angular structures simulated include top hat, multiple power-law structures,  Gaussian jets and  hollow jets (see structure definitions in Table \ref{tab: sim_setup}).
Given an angular energy structure $E_{iso}(\theta)$, we set the initial conditions at each angle as part of a BM solution with the local value of $E_{iso}(\theta)$. Additional information on the simulation setup can be found in Appendix \ref{appendix:sim_setup}.

\begin{table*}
    \centering
\begin{tabular}{|c|c|c|c|c|}
\hline 
\multicolumn{2}{|c|}{Simulation} & Initial $\frac{dE}{d\Omega}\propto$ & $\theta_{i}$ & $\gamma_{i}$\tabularnewline
\hline 
\hline 
\multicolumn{1}{|c}{\multirow{7}{*}{top hat}} & \multirow{7}{*}{} & \multirow{7}{*}{$\begin{cases}
1 & \theta\le\theta_{i}\\
0 & \theta>\theta_{i}
\end{cases}$} & $0.005$ & 400\tabularnewline
\cline{4-5} \cline{5-5} 
 &  &  & $0.02$ & 200\tabularnewline
\cline{4-5} \cline{5-5} 
 &  &  & $0.05$ & 100\tabularnewline
\cline{4-5} \cline{5-5} 
 &  &  & $0.1$ & 100\tabularnewline
\cline{4-5} \cline{5-5} 
 &  &  & $0.15$ & 100\tabularnewline
\cline{4-5} \cline{5-5} 
 &  &  & $0.2$ & 100\tabularnewline
\cline{4-5} \cline{5-5} 
 &  &  & $0.3$ & 100\tabularnewline
\hline 
\multirow{9}{*}{Power-Law} & $b=2.2$ & \multirow{9}{*}{$\propto\begin{cases}
1 & \theta\le\theta_{i}\\
\left(\frac{\theta}{\theta_{i}}\right)^{-b} & \theta>\theta_{i}
\end{cases}$} & $0.02$ & 80\tabularnewline
\cline{2-2} \cline{4-5} \cline{5-5} 
 & $b=2.5$ &  & $0.02$ & 200\tabularnewline
\cline{2-2} \cline{4-5} \cline{5-5} 
 & \multirow{2}{*}{$b=3$} &  & $0.05$ & 100\tabularnewline
\cline{4-5} \cline{5-5} 
 &  &  & $0.15$ & 100\tabularnewline
\cline{2-2} \cline{4-5} \cline{5-5} 
 & \multirow{3}{*}{$b=6$} &  & $0.02$ & 200\tabularnewline
\cline{4-5} \cline{5-5} 
 &  &  & $0.05$ & 100\tabularnewline
\cline{4-5} \cline{5-5} 
 &  &  & $0.15$ & 100\tabularnewline
\cline{2-2} \cline{4-5} \cline{5-5} 
 & \multirow{2}{*}{$b=12$} &  & $0.05$ & 100\tabularnewline
\cline{4-5} \cline{5-5} 
 &  &  & $0.15$ & 100\tabularnewline
\hline 
$b6c1$ & $b=6$ & $\propto\begin{cases}
1 & \theta\le\frac{\theta_{i}}{2}\\
\left(\frac{\theta}{0.5\theta_{i}}\right)^{-1} & \frac{\theta_{i}}{2}<\theta\le\theta_{i}\\
0.5\left(\frac{\theta}{\theta_{i}}\right)^{-b} & \theta>\theta_{i}
\end{cases}$ & $0.02$ & 200\tabularnewline
\hline 
\multirow{2}{*}{Gaussian} &  & \multirow{2}{*}{$\propto\exp\left[-\left(\frac{\theta}{\theta_{i}}\right)^{2}\right]$} & $0.02$ & 200\tabularnewline
\cline{4-5} \cline{5-5} 
 &  &  & $0.07$ & 100\tabularnewline
\hline 
\multirow{2}{*}{Hollow} & \multicolumn{2}{c|}{$\propto\begin{cases}
\frac{1}{1+\exp\left[-5\left(\frac{\theta}{\theta_{i}}-\frac{1}{2}\right)\right]} & \theta\le\frac{\theta_{i}}{2}\\
1 & \frac{\theta_{i}}{2}<\theta\le\theta_{i}\\
0 & \theta>\theta_{i}
\end{cases}$} & $0.05$ & 100\tabularnewline
\cline{2-5} \cline{3-5} \cline{4-5} \cline{5-5} 
 & $b=6$ & $\propto\begin{cases}
\frac{\theta}{\theta_{i}} & \theta\le\frac{\theta_{i}}{2}\\
1 & \frac{\theta_{i}}{2}<\theta\le\theta_{c}\\
\left(\frac{\theta}{\theta_{i}}\right)^{-b} & \theta>\theta_{i}
\end{cases}$ & $0.15$ & 100\tabularnewline
\hline 
\end{tabular}
    \caption{Initial conditions for simulations used in this work. $\gamma_i$ denotes the highest matter Lorentz factor right behind the shock in the initial conditions.}
\end{table*}\label{tab: sim_setup}

\subsection{Evolution of top hat jets}\label{sec:th}
We start by considering the evolution of top hat jets, before moving on to extend our study to more general structures. Before validating and calibrating the analytic equations for the evolution of $\Gba$ and $\theta_c$, we briefly discuss the initial conditions of top hat jets.
The exact initial conditions of top hat jets in our simulations are not expected in nature, and they are "unstable" in the sense that the hydrodynamic evolution changes the blast wave structure over a duration that is much shorter than the dynamical timescale of the system. Thus, following the beginning of the simulation the blast wave quickly assumes a more "stable" structure (i.e., one that evolves over a dynamical timescale) via a flow of energy outside of the core.  The result is that during the pre-spreading phase, the core angle of the blast wave, $\theta_c$, is smaller than $\theta_i$, the initial opening angle of a top hat jet. In the following we approximate the core angle during the pre-spreading phase as constant and we denote this angle by $\theta_{c,0}$.

We can now go on to validate and calibrate the equations from \S\ref{sec:analytical}, as well as the value of $\qco$, and then we go on to investigate the jet structure evolution during the pre-spreading phase and the spreading self-similar evolution. 

\subsubsection{Evolution of $\Gba$ and $\theta_c$}

We adopt the following definitions of $\qc$  from \cite{GovreenSegal2023}: 
\begin{equation}\label{eq:qc_def}
    \theta_c \equiv \theta\left(\frac{d\log E_{iso}}{d\log \theta}=-2\right).
\end{equation}
We choose this definition for the following reasons. First, it can be identified from the observations, since it corresponds to the region of the jet that dominates the emission at the time of the peak of the light curve of off-axis jets \citep{GovreenSegal2023}. Second, this definition is also meaningful in terms of jet energy. For a jet structure in which $\frac{d \log E_{iso}}{d\log \theta}$ decreases monotonously, this definition means that in the core the energy per $\Delta \theta \sim \theta$ increases with $\theta$, and outside the core the added energy decreases with the angle. This promises that the jet core dominates (or at least contains a significant fraction of) the total jet energy. In our simulations, we find that while this definition works very well during the relativistic phase it cannot be extended to the mildly relativistic phase. The reason is that $\frac{d \log E_{iso}}{d\log \theta}$ evolves during the transition to the sub-relativistic phase in a way that the location of $\theta_c$ according to this definition is not continuous. See appendix \ref{appendix: theta_c} for a comparison of various definitions of $\theta_c$.

\begin{figure}
    \centering    \includegraphics[width=\columnwidth]{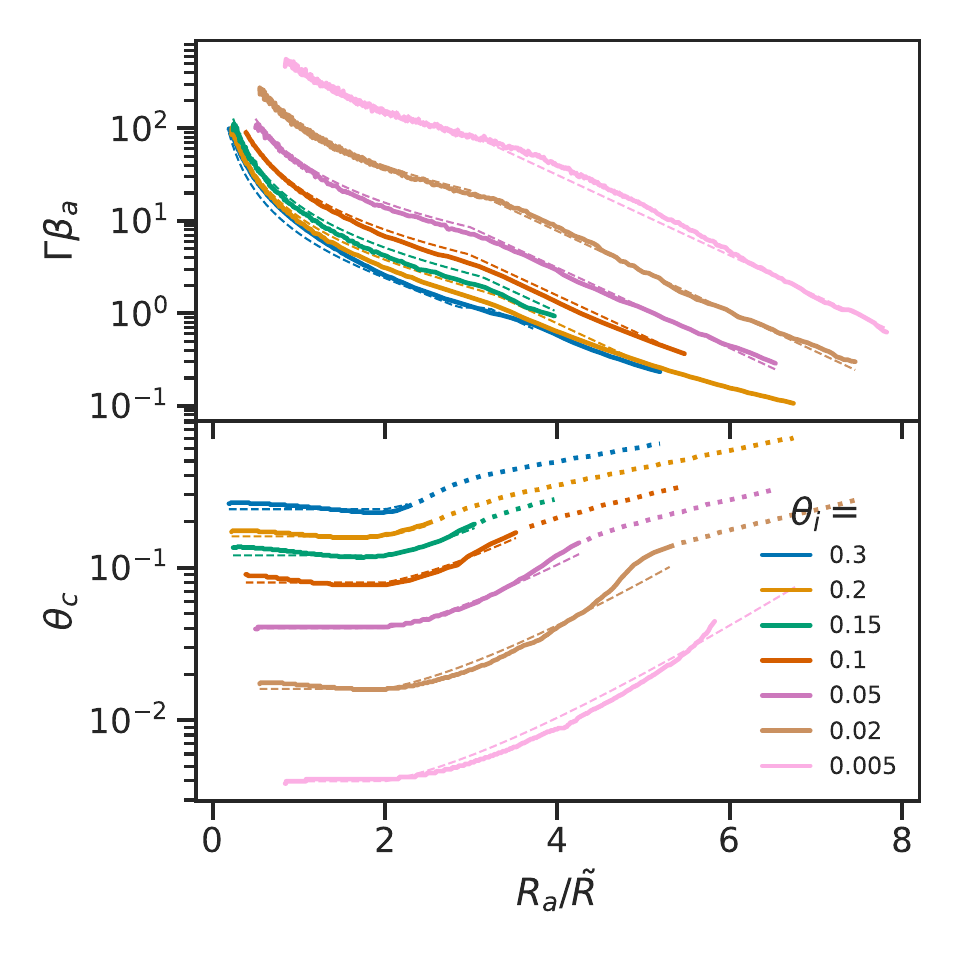}
    \caption{The evolution of the proper velocity along the axis, $\Gba$ ({\it top panel}), and core angle, $\qc$ ({\it bottom panel}), of top hat jets with various opening angles. The plot is semi-log with radius normalized by $\Rt$, to elucidate the exponential dependence during the spreading phase. The simulation curves are plotted in solid lines, with exception of the angles in the bottom panel that are plotted as dots after the shock becomes mildly relativistic ($\Gba<2$). The analytic approximation is plotted in dashed lines - Eqs. \eqref{eq:Gbm} and \eqref{eq:Gsp} for $\Gba$ and Eqs. \eqref{eq:qc} and \eqref{eq:q0} for $\qc$.}
    \label{fig:U_a,th_c}
\end{figure}

Fig. \ref{fig:U_a,th_c} shows a comparison of our analytic model with  $\tilde{C}=0.65$, $C_{tot}=0.54$, and $C_{sp}=0.38$ to the results of all our simulations of top hat jets.  Both $\Gba$ and $\theta_c$ fit the analytical model well as long as the jet is relativistic, and reasonably well while the jet is mildly relativistic, for initial opening angles up to $0.3$ rad. 

Note that $\theta_c$ starts spreading at $R\simeq 2\Rt$ almost at half the radius at which the proper velocity along the axis starts dropping exponentially, $\Rs$, this is because they are both local properties and the edge of the core is affected by the spreading before the jet axis. Using this, we can define $\qco$:
\begin{equation}\label{eq:q0}
    \qco = \qc(R=2\tilde{R}) \simeq 0.5 \qsp
\end{equation}
which implies $\qco \approx 0.8 \theta_i$. This is indeed smaller than the top hat jet initial opening angle, as we expected. These results imply that $\Gbs  \simeq 0.24 \theta_i^{-1} \simeq 0.19 \theta_{c,0}^{-1}$, which explains part of the difficulty to observe the relativistic spreading phase in previous simulations. For example, in a simulation of a top hat jet with $\theta_i=0.1$ rad, the proper velocity along the axis at $\Rs$ is $\Gbs \simeq 2.4$, so the shock is already mildly relativistic at the begining of the spreading phase.
\begin{figure}
    \centering
    \includegraphics[width=\columnwidth]{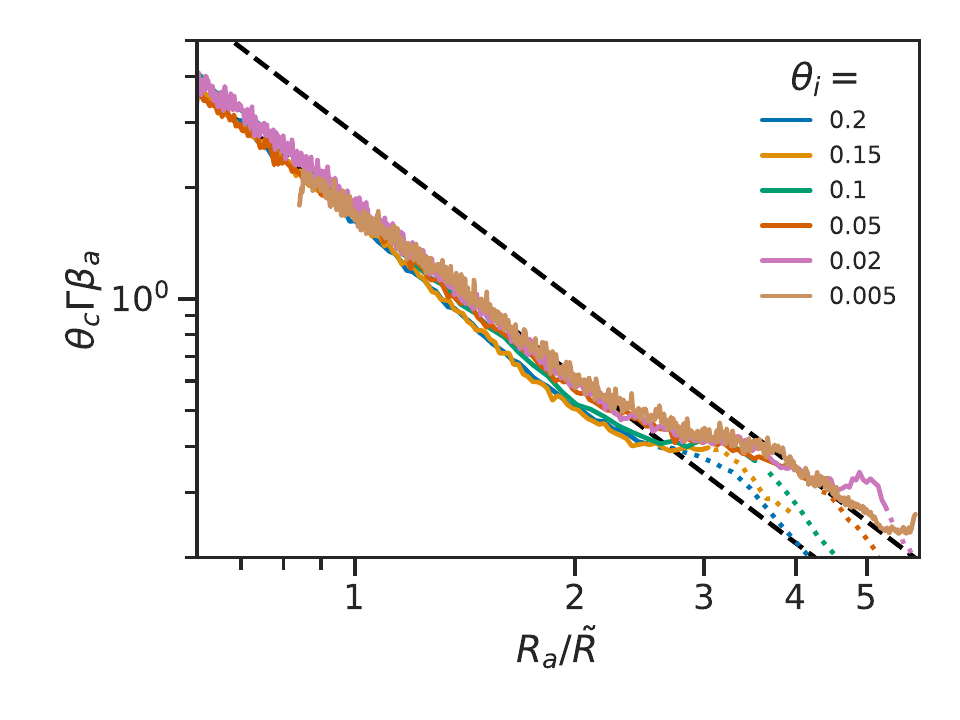}
    \caption{The product of the core angle and proper velocity along the axis as a function of $R_a/\Rt$. Both in the BM phase and in the spreading phase, the product is proportional to $R^{-3/2}$ (marked in black dashed lines), however, the normalization is different. Of all the top hat jet simulations, only the one with $\theta_i=0.005$ has enough dynamical range to clearly show this scaling in the spreading phase. The simulation curves are plotted in solid lines in the relativistic regime ($\Gba>2$) and as dots after the shock becomes mildly relativistic ($\Gba<2$).}
    \label{fig:gba_thc}
\end{figure}

Figure \ref{fig:gba_thc} shows the product $\Gba\qc$ as a function of $R_a/\Rt$. In both the pre- and post-spreading relativistic regimes energy conservation dictates $\Gba\qc \propto R_a^{3/2}$. However, the normalization, which reflects the ratio between the core energy and the total energy is expected to be different between the top hat pre-spreading phase and the self-similar post-spreading phase. In Figure \ref{fig:gba_thc} 
we see that all top hat jets follow the same track in this relation while the shock is relativistic, but only the simulation with $\theta_i=0.005$ starts the spreading phase relativistic enough to follow the relativistic spreading curve of $\Gba\qc\propto R^{-3/2}$ for a significant period before becoming mildly relativistic. Note, that this figure validates the energy conservation equation \ref{eq:Ec} and the fact that the collected mass behind the core satisfies the "conical" relation $M \propto R^3 \theta_c^2$.

\subsubsection{Pre-spreading jet structure}\label{sec:pre-spreading} 
\begin{figure}
    \centering
\includegraphics[width=\columnwidth]{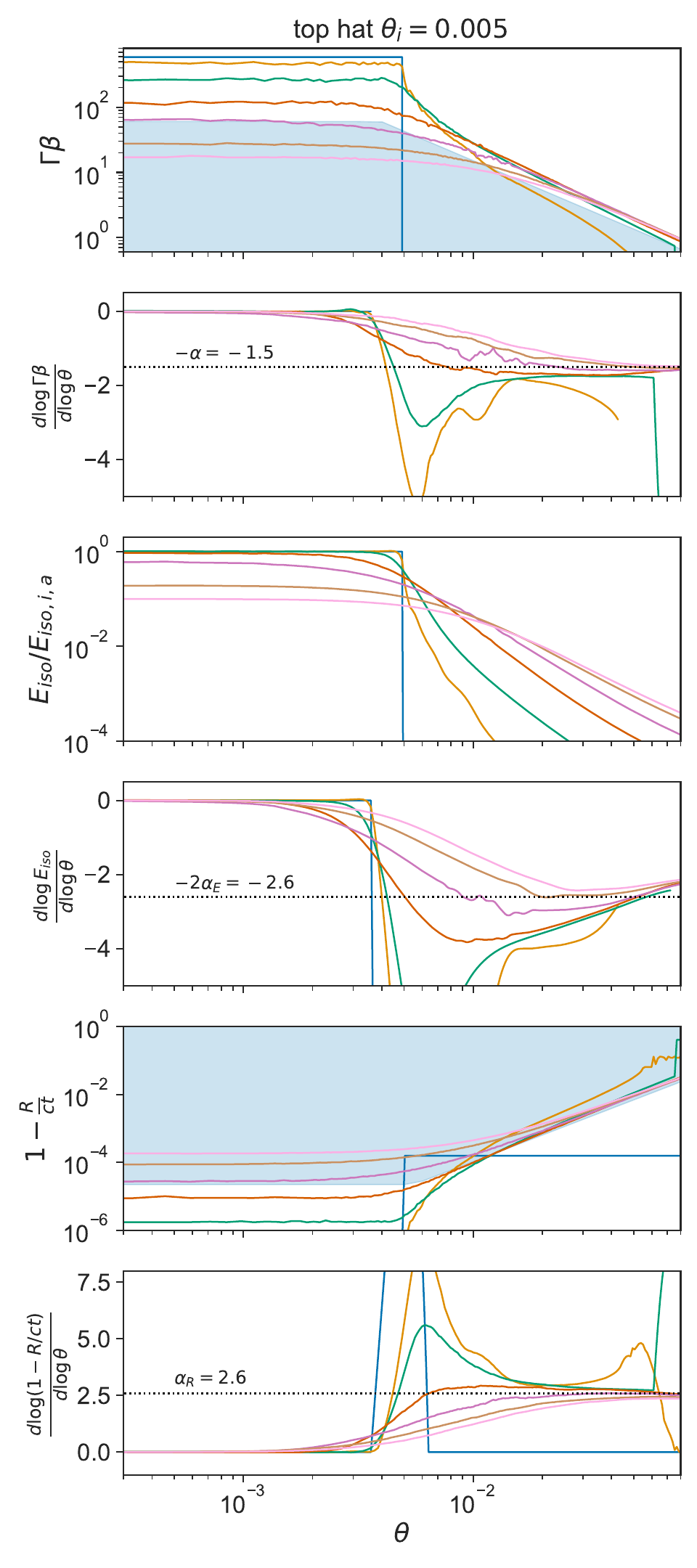}
    \caption{The evolution of $\Gamma\beta(\theta)$, $E_{iso} (\theta)$ and $(1-\frac{R(\theta)}{ct})$ is plotted for a top hat jet with an initial opening angle of $0.005$ rad. {\it Top panel}:  $\Gamma\beta(\theta)$. Lower curves correspond to later times. The "wings" of the structure approach the edge of the shaded region and move with a constant velocity. {\it Second panel}: The logarithmic derivative of $\Gamma\beta$ with respect to $\theta$, the wings  approach a power-law of $\simeq -1.5$. {\it Third panel}: $E_{iso} (\theta)$. During the pre-spreading phase energy flows continuously from the core to the wings, which approach the self-similar solution. {\it Fourth panel}: The logarithmic derivative of $E_{iso}$ with respect to $\theta$, the wings can be seen approaching a power-law of $\simeq -2.6$. {\it Fifth panel}: The normalized distance between the shock radius and $ct$ as a function of angle. {\it Sixth panel}: The logarithmic derivative of $1-\frac{R}{ct}$ with respect to $\theta$, the wings approach a power-law of $\simeq -2.6$.}
    \label{fig:tophat_0.005}
\end{figure}

We use the simulation of a top hat jet with an initial opening angle of $0.005$ rad, and an initial shock Lorentz factor of $\Gamma_i \simeq 560$ (the initial Lorentz factor of the matter just behind the shock is $\gamma_i=400$) to investigate the pre-spreading evolution, as this is the simulation with the most dynamical range. Fig. \ref{fig:tophat_0.005} portrays $\Gb(\theta),E_{iso}(\theta)$ and $(1-\frac{R}{ct})(\theta)$  at various times, as well as the logarithmic derivatives of these quantities with respect to $\theta$. 
First, as expected, it shows two phases. A {\it pre-spreading phase}, during which the core angle is almost constant and $E_{iso}$ along the jet axis does not change in time, and a {\it spreading phase} during which the core angle increases significantly, and energy is transferred to larger angles. The transition between the two is when the shock Lorentz factor is lower by a factor of a few compared to one over the initial opening angle.  A second interesting feature is that the blast wave develops wings outside of the core immediately after the beginning of the simulation (already during the pre-spreading phase). These wings exhibit an interesting behavior: in each of the properties plotted, the wings approach a constant profile. $\Gamma\beta \propto \theta^{-\alpha}$ where $\alpha \simeq 1.5$, $E_{iso} \propto \theta^{-2\alpha_E}$, where $\alpha_E \simeq 1.3$, and  $1-\frac{R}{ct}\propto\theta^{\alpha_R}$, where $\alpha_R \simeq 2.6$. 
Note, that unlike the BM solution and the conditions at the core where $\Gamma \propto E_{iso}^{1/2}$, in the wings $\alpha_E \neq \alpha$ and $\Gamma \not \propto E_{iso}^{1/2}$. This is evidence for the continuous lateral transfer of energy through the wings. The value $\alpha_R$ similarly provides evidence for the lateral motion of matter. If the radial location of the matter was strictly due to its own radial velocity, we would expect a power-law of $3$ or steeper, rather than $2.6$.

We can fit the numerical pre-spreading evolution of a top hat jet structure. The proper velocity is roughly uniform in the jet core and is therefore roughly $\Gba$. The wings develop a power-law, with a constant index where at each angle the shock propagate at a constant velocity. Altogether, We find the structure is well described by:
\begin{equation}\label{eq:Gb_bs}
    \Gamma \beta \simeq
    \begin{cases}
\Gba & \theta <\theta_c\\
\frac{0.38}{\theta_{c}}\left(\frac{\theta}{\theta_{c}}\right)^{-\alpha} & \theta >\theta_c
\end{cases}
   ~~~R>\Rs ~~~,
\end{equation}
where $\alpha\simeq1.5$. 

At the same time, the energy profile is constant in the core and it develops wings outside of the core, which gradually approach $E_{iso}\propto\left(\frac{\theta}{\theta_c}\right)^{-2\alpha_E}$.
Like the energy and the Lorentz factor, the shock radius is also uniform in the core, and can be easily found by integrating over the shock velocity in the core. We use this to determine the normalization of $(1-\frac{R}{ct})$, and find: 
\begin{equation}
    1-\frac{R}{ct} \simeq
    \begin{cases}
\frac{1}{8\Gba^2} & \theta <\theta_c\\
0.6\cdot \theta_c^2\left(\frac{\theta}{\theta_c}\right)^{\alpha_R} & \theta >\theta_c
\end{cases}
   ~~~R>\Rs ~~~,
\end{equation}
where $\alpha_R=2.6$.
 
\subsubsection{Spreading phase - self-similar Evolution}\label{sec:post-spreading}
\begin{figure}
    \centering
    \includegraphics[width=\columnwidth]{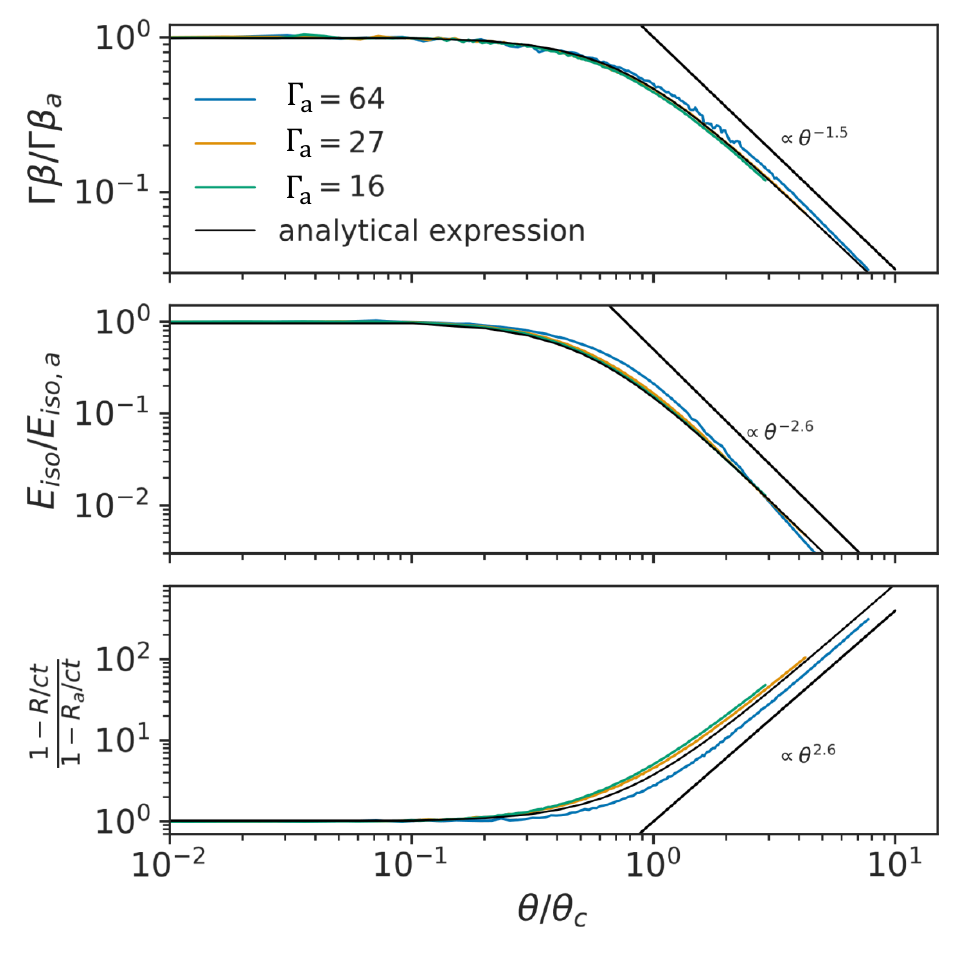}
    \caption{The bottom three curves from the top panel of Fig. \ref{fig:tophat_0.005}, plotted in normalized coordinates. The blue curve is from the final stages of the transition into the self-similar solution, while the two later jet profiles indicate a self-similar evolution. The analytic fits to smoothed broken power-laws (Eqs. \ref{eq:U_self-simiar}-\ref{eq:Eiso_self-simiar}) are marked in black.}
    \label{fig:tophat_self-similar}
\end{figure}
After the spreading starts we find that the blast wave assumes what seems as a self-similar lateral structure\footnote{Numerically we cannot prove that the structure is indeed fully self-similar. Nevertheless, given the agreement of the evolution with the self-similar scaling \citep{Gruzinov2007,Keshet2015} and that all our tests agree with a self-similar evolution, we assume throughout our discussion that it is indeed self-similar. }. It is shown in Fig. \ref{fig:tophat_self-similar}, which depicts the angular structure at various times in normalized dimensions; $\Gamma\beta/\Gba$, $E_{iso}/E_{iso,a}$ and $(1-R/ct)/(1-R_a/ct)$ as a function of $\theta/\theta_c$. This figure also shows analytic smoothed broken power-law approximations for these structures. The approximation for the proper velocity is:
\begin{equation}\label{eq:U_self-simiar}
    \Gb \simeq \Gba \left(1+\left(\frac{\theta}{k\theta_c}\right)^{1/s}\right)^{-s \cdot \alpha }=\Gba \left(1+\left(\frac{4\theta}{3\theta_c}\right)^{2}\right)^{-0.75 }~, 
\end{equation}
where in the second equality we set $\alpha=1.5$ and the calibration coefficients $s=0.5$,  $k=0.75$ (corresponding to $\Gb(\theta_c)=0.46\Gba$). 
Similarly, the radius follows a self-similar structure:
\begin{equation}\label{eq:R_self-similar}
\begin{split}
        1-\frac{R}{ct}&=\left(1-\frac{R_a}{ct}\right)\left(1+\left(\frac{\theta}{k \theta_c}\right)^{1/s_R}\right)^{\alpha_R s_R}\\&=\left(1-\frac{R_a}{ct}\right)\left(1+\left(\frac{4\theta}{3 \theta_c}\right)^2\right)^{1.3}.
\end{split}
\end{equation}
where we find $s_R=0.5$ and $\alpha_{R}=2.6$. 
The self-similar structure of $E_{iso}$ can also be approximated very well by a smoothed broken power law:
\begin{equation}\label{eq:Eiso_self-simiar}
    E_{iso} \simeq E_{iso,a} \left(1+\frac{1}{\alpha_{E}-1}\left(\frac{\theta}{\theta_c}\right)^{\frac{1}{s_E}}\right)^{-2\alpha_E s_E}= E_{iso,a}\left(1+\frac{3.3\theta^2}{\theta_c^2}\right)^{-1.3}, 
\end{equation}
where $\alpha_E=1.3$ and $s_E=0.5$. The $\frac{1}{\alpha_E-1}$ factor ensures that $\frac{d\log E_{iso}}{d\log\theta}\left|_{\theta_c}\right.=-2$. Note, however, that $E_{iso}$ takes longer to approach the self-similar solution, most probably because it involves integrating over the complete radial coordinate and is, therefore, more sensitive to the evolution history, while $\Gb$ and $R$ can change more rapidly. 

Another interesting aspect of this phase is the hydrodynamic radial profile. We find that the pressure, density and velocity all have what seems as a self similar radial structure both in time (i.e., at a given angle at different times) and in angle (i.e., at a given time at different angles). We discuss this in Appendix \ref{appendix: GSS}.

The self-similar structure and its evolution enable us to define an important angle - $\qNR$, the location along the wings where the blast wave becomes sub-relativistic ($\Gb(\qNR)=1$) during the spreading phase. Using the wing structure, $\Gb(\theta) \propto \theta^{-\alpha}$, we can express this angle as $\qNR = \qc \Gb_{\theta=\qc}^{1/\alpha}$, and using equations \eqref{eq:Gsp}, \eqref{eq:qc} and \eqref{eq:U_self-simiar} we obtain:
\begin{equation}
        \qNR = 0.31 \qsp^{1/3} \left(\frac{R_a}{\Rs}\right)^{-1.5}\exp\left(\frac{1}{3}\frac{R_a-\Rs}{\Rt}\right).
\end{equation}
While asymptotically $\qNR$ grows exponentially with the radius, in the range relevant to us, this function varies very little before the jet core becomes mildly relativistic. Qualitatively, $\qNR$ initially decreases as a power-law in $R$, and only then starts increasing. At the onset of spreading, at $\frac{R}{\Rt}=3.6$, $\qNR = 0.315 \qsp^{1/3}$, while at it's minimum, $\frac{R}{\Rt}=4.5$ and $\qNR = 0.305 \qsp^{1/3}$. For example, a jet with $\qsp = 0.1$ rad ($\qco\simeq0. 05$ rad), becomes mildly relativistic at $\frac{R}{\Rt}=4.9$, when it opening angle is only $0.306\qsp^{1/3}$. A jet with $\qsp=0.01$ rad ($\qco\simeq0.005$ rad) will become mildly relativistic at $\frac{R}{\Rt}=7.2$ when $\qc \simeq 0.37\qsp^{1/3}$. Even a jet with unrealistic $\qsp=0.001$ rad ($\qco\simeq0.0005$ rad), which becomes mildly relativistic at $\frac{R}{\Rt}=9.5$, have at this time an opening angle of $0.52\qsp^{1/3}$, meaning that even in this extreme case, where the core angle grows by two orders of magnitude during the relativistic phase (from 0.0005 to 0.05 rad), the maximal angle at which there is relativistic matter,$\qNR$, grows by less than a factor of 2 (from 0.03 to 0.05 rad). In the simulations, we find that $\qNR$ is approximately constant not only during the spreading phase, but also during the entire pre-spreading phase of top hat jets (see Fig. \ref{fig:tophat_0.005}). 
We conclude that for any reasonable jet, we can use:
\begin{equation}\label{eq:qNR}
    \qNR\simeq0.31\qsp^{1/3} \simeq 0.39\qco^{1/3}.
\end{equation}
The fact that $\qNR$ is nearly-constant, is consistent with the fact that at any given angle along the jet wings, the Lorentz factor remains roughly constant throughout the self-similar evolution, and also earlier as the jet approaches the self-similar structure. 

Eq. \ref{eq:qNR} has an interesting implication. Although the jet expands significantly during the spreading phase, it remains highly collimated even at the time that the shock becomes mildly relativistic. For example, for $\qco=0.05$ rad (as observations suggest for the jet in GW 170817) the opening angle when $\Gba=1$ is only $0.14$ rad (about $8^\circ$). In addition, the opening angle at the transition to the sub-relativistic regime depends on the initial opening angle, and it therefore has a "memory" of the initial conditions.  This is in contrast to the naive expectation that $\Gamma \theta_c \sim 1$ (which implies $\qNR \sim 1$ rad) and inline with previous 
numerical results \citep{vanEerten2013,Duffell2018}. $\qNR \ll 1$ mainly due to the fact that while $\Ga \propto \exp [-R_a/\Rt]$, the exponential spreading of $\qc$ is mitigated by a power-law term ($\qc \propto R_a^{-3/2}\exp [R_a/\Rt]$). Thus, the core angle spreads "slower" than the shock decelerates and it remains highly collimated also when the shock becomes sub-relativistic.

\subsection{Evolution of a jet with a general structure}\label{sec:structured}
Next, we generalize our findings for top hat jets to jets with a general structure. 
In the top hat case, we have seen that the jet develops wings that approach self-similar energy and Lorentz factor gradients, passing through a range of increasingly shallower jet structures in the process. We, therefore, expect any jet with an initial structure in which the wings are steeper than in the self-similar solution to approach the self-similar solution. While we don't have an exact definition of which jets will approach the self-similar solution, we expect a self-similar structure to form for all jets in which $E_{iso,i}(\theta)$ in the wings decreases more sharply than $\simeq \theta^{-3}$, though the convergence to the self-similar structure depends on the exact initial structure. Indeed, we find that the $b\ge3$ power-law jets, Gaussian jets, and even our hollow jet simulation all approach the same self-similar solution as top hat jets.

The top hat results, however, do not indicate what will be the evolution of a jet with wings that are wider than those of the self-similar solution. Therefore, we separate our discussion of general jet structures into two cases. First, we discuss the case of a jet with initial wings that are steeper than the self-similar profile. We denote this case as a "narrow structure". The second case is jets with an initial structure that is wider than the self-similar solution, which we denote as a "wide structure".

\subsubsection{Evolution of "narrow structure" jets}
\begin{figure}
    \centering
\includegraphics[width=\columnwidth]{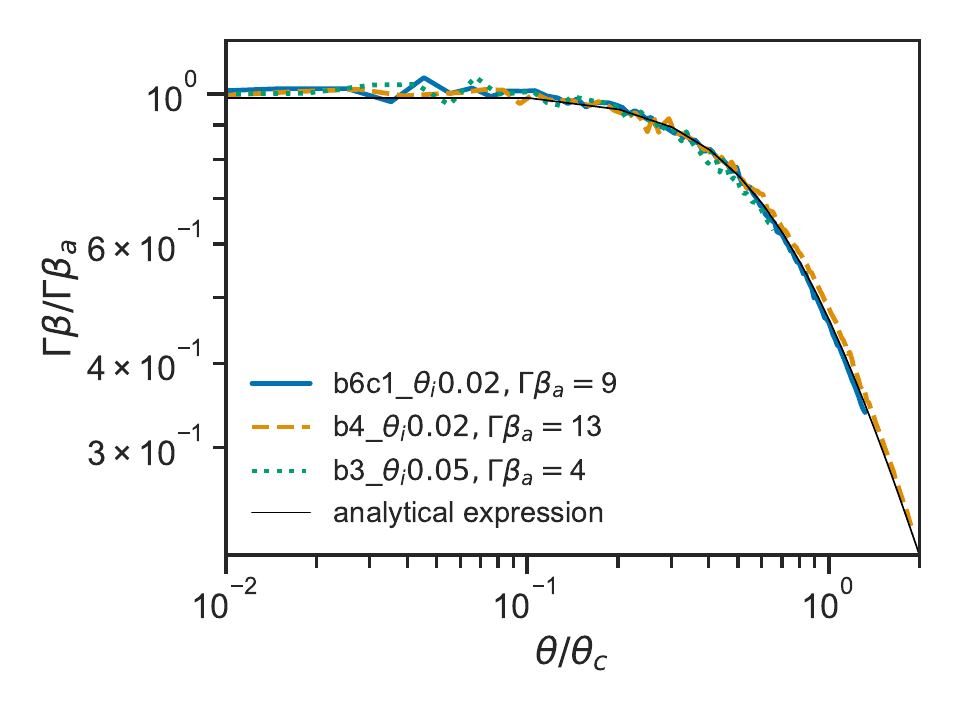}
    \caption{The normalized shock proper-velocity profile in the self-similar regimes. Various simulations and times coincide with a similar structure. The thin black solid line is the smoothed broken power-law fit (Eq. \ref{eq:U_self-simiar})}
    \label{fig: self-similar structure}
\end{figure}

As we saw in \S\ref{sec:th}, the self-similar solution to which a top hat jet evolves can be fully defined in terms of two parameters. The first defines the natural length scale of the jet evolution, $\Rt$, which can be found from the initial condition $E_{tot}/\rho$ (Eq. \ref{eq:Rt}). The second defines the angular scale of the solution, $\qsp$, which can be found from the initial condition $\frac{E_{tot}}{E_{iso,i}}$ (Eq. \ref{eq:qsp}).
We expect that all jets with narrow structures will evolve to the same self-similar structure found in top hat jets, and indeed all our simulations of narrow jets do (a few examples are shown in Fig. \ref{fig: self-similar structure}). The main question is then, what is the mapping from the jet initial structure,  $E_{iso,i}(\theta)$, to the length scale $\Rt$ and the angle $\qsp$ of the self-similar solution that it assumes. We did not find functions that provide an exact mapping of a general initial structure to the self-similar solution that it evolves to, and it is poossible that there are no such simple and general functions. Instead we find a simple approximation that agrees well with our set of simulations. According to this approximation, in  Eqs. \eqref{eq:Gbm}-\eqref{eq:Rsp}, $E_{tot}$ is taken as the total jet energy, including the wings, $E_{tot}=\intop\frac{dE}{d\Omega} d\Omega$, and $E_{iso,i}$ is replaced with the energy weighted average initial isotropic equivalent energy as defined by \cite{Duffell2011},
\begin{equation}\label{eq: Eiso}
    \Eiso=4\pi \frac{\intop \left(\frac{dE_i}{d\Omega}\right)^2 d\Omega}{\intop\frac{dE_i}{d\Omega} d\Omega}.
\end{equation}
Note that for a top hat jet, this definition coincides with $E_{iso,i}$. The three  order unity parameters that were found numerically, $\tilde{C}$, $C_{tot}$ and $C_{sp}$, are the same as for the case of top hat jets, since they reflect the properties of the self-similar solution.   

The core angle before the onset of spreading is more difficult to approximate, as it depends on the transition between the initial structure and the self-similar solution, during which the core angle evolves in some manner from $\theta_i$, to $\qco$, but this manner is not necessarily monotonic. Approximating $\theta_c\simeq\qco$ (where $\qco$ is taken from Eq. \ref{eq:q0})  before the onset of spreading is a reasonable choice, although it can result in errors of up to a factor of 2. 
\begin{figure}
    \centering    
    \includegraphics[width=\columnwidth]{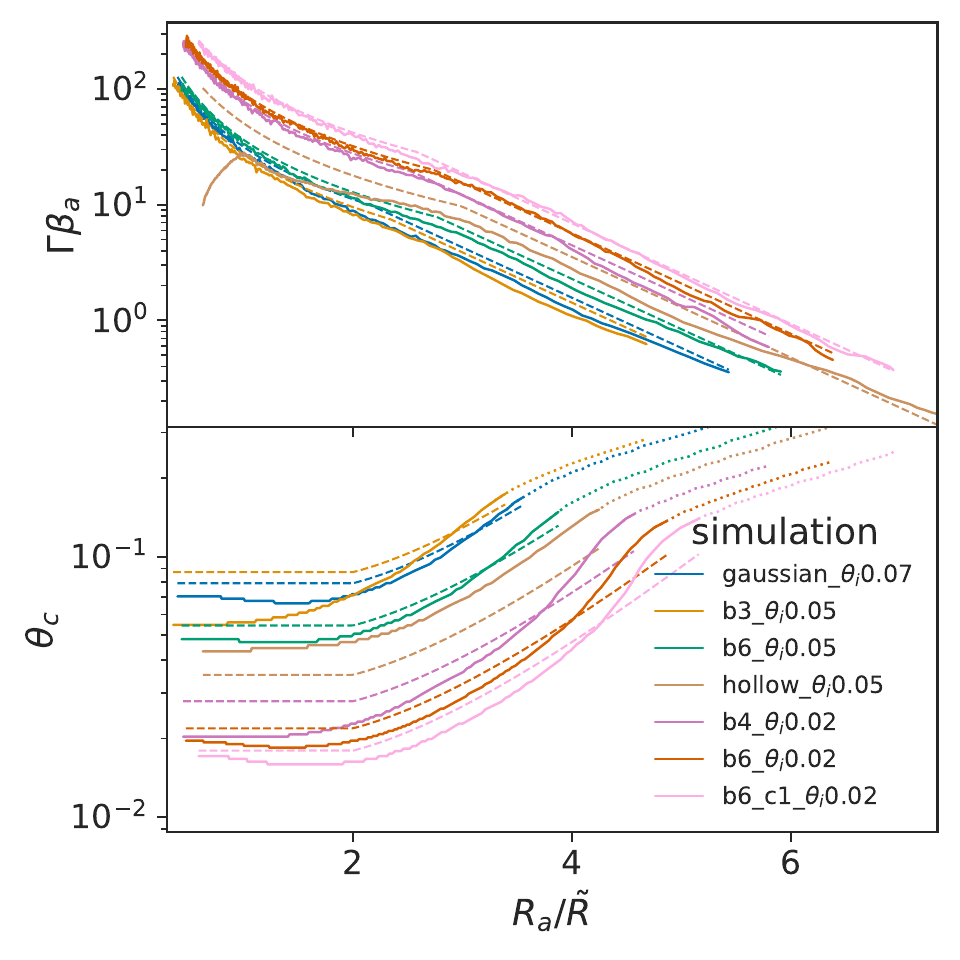}
    \caption{Same as Fig, \ref{fig:U_a,th_c}, for a variety of jet structures.}
    \label{fig:U_a,th_c_all}
\end{figure}
The quality of the approximation can be seen in Fig. \ref{fig:U_a,th_c_all}.
Note that for jets in which the energy profile varies significantly within the core, such as hollow jets (the red curve), the initial phase in which the core flattens out is not included in our model, and using $\Eiso$ is a reasonable choice only from when the core becomes approximately flat. Even then, the approximation is not nearly as good as the description we find for other jets.

\begin{figure*}
    \centering
    \includegraphics[width=\textwidth]{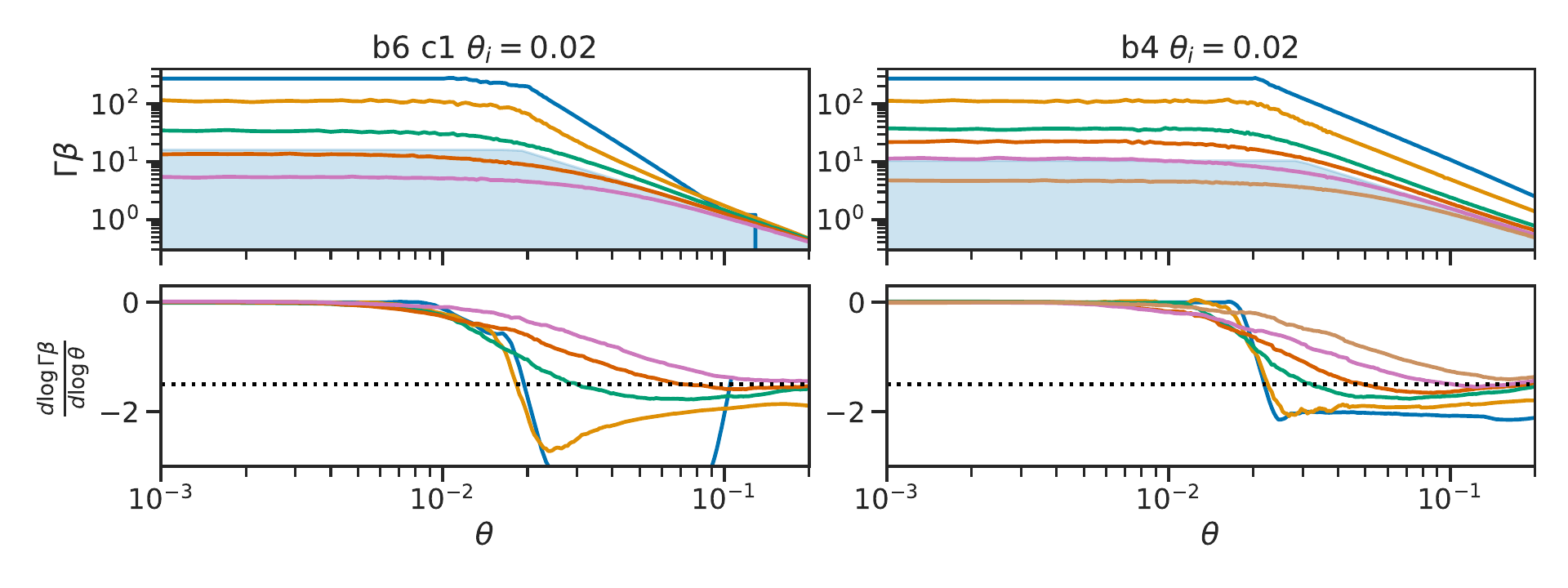}
    \caption{The evolution of $\Gb(\theta)$ and $\frac{d\log\Gb}{d\log\theta}$ is plotted for two structured jets. In the two {\it left} panels, simulation b6c1 (see Table \ref{tab: sim_setup}) - a jet with the following core structure:  $E_{iso,i}(\theta\le \theta_{i}/2)=const$  and $E_{iso,i}(\theta_{i}/2\le \theta\le\theta_{i}) \propto\theta^{-1}$. Outside the core $E_{iso,i}(\theta>\theta_i)\propto\theta^{-6}$. In the {\it right} panels, a $b=4$ power-law jet with a flat core. Both jets show the expected evolution, where the wings approach a power law with an index $-\alpha$. During the spreading phase both jets assume the same self-similar profile as top hat jets (marked by the shaded region), and during this phase their evolution is generic and  independent of the initial jet structure.}
    \label{fig:b6_b4_structure}
\end{figure*}

The evolution of the jet structure depends on its initial angular profile and the initial Lorentz factor. Approximating $\qNR$ as constant, we can qualitatively describe it as follows. If initially $\theta(\Gb=1)<\qNR$, the structure will quickly widen to fill the region upto $\qNR$ with relativistic material. Then, as the core decelerates, the profile is gradually approaching the power law $\Gb\propto\theta^{-\alpha}$, reaching it at the onset of the self-similar solution. 
If initially $\theta(\Gb=1)>\qNR$, then at every angle, the structure will initially evolve as part of a BM solution with the local value of $E_{iso,i}(\theta)$, until it decelerates enough that $\theta(\Gb=1)=\qNR$. From this point, the structure will start flattening out and approaching the self-similar solution.

Fig.  \ref{fig:b6_b4_structure} shows two examples for the structure evolution of power-law jets  before the onset of spreading, both of "narrow structure" jets. In one (b=6, left panels) initially $\theta(\Gb=1)<\qNR$ and in the other (b=4, right panels) initially $\theta(\Gb=1)>\qNR$. In both cases, the jets approach the self-similar solution in the manner described above.
While we do not find an exact formula for the structure evolution, we find that once $\theta(\Gamma\beta=1)=\qNR$, the structure of $\Gb$ in our simulations evolves approximately as a single power-law spanning from $\qc$ to $\qNR$:
\begin{equation}\label{eq:Gb_bs}
    \Gamma \beta \simeq
    \begin{cases}
\Gba & \theta <\theta_c\\
\left(\frac{\theta}{\qNR}\right)^{-\alpha_{eff}} & \theta >\theta_c
\end{cases}
   ~~~R>\Rs ~~~,
\end{equation}
where approximating $\Gamma\beta_c\simeq\Gba$, $\alpha_{eff}=-\frac{\log \Gba}{\log\frac{\qc}{\qNR}}$. Similarly:
\begin{equation}
    1-\frac{R}{ct} \simeq
    \begin{cases}
\frac{1}{8\Gba^2} & \theta <\theta_c\\
\left(1-\frac{R_a}{ct}\right)\cdot\left(\frac{\Gb}{\Gba}\right)^{-1.73} & \theta >\theta_c
\end{cases}
   ~~~R>\Rs ~~~,
\end{equation}
that is, the relation between $1-\frac{R}{ct}$ and $\Gb$ remains as in the self-similar case, and here too the power-law gradually changes. Using these approximations provides a reasonable approximation for the light curve. Formulae that can be used for approximation of power-law jets as well as several examples of light curves are shown in Appendix \ref{appendix:structure_approx}.

\begin{figure}
    \centering    \includegraphics[width=\columnwidth]{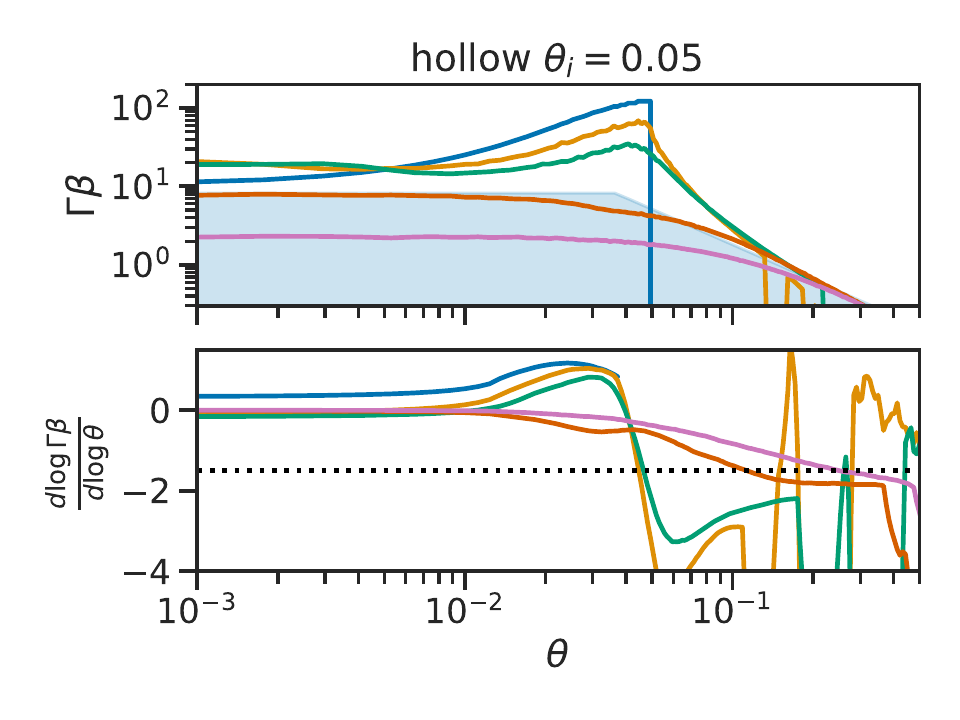}
    \caption{The evolution of $\Gb(\theta)$ and $\frac{d\log\Gb}{d\log\theta}$ for a hollow jet. The core is flattened significantly during the pre-spreading phase, and it  approaches the self-similar structure (plotted schematically in shaded blue) by the transition to the spreading phase.}
    \label{fig:hollow_U_theta}
\end{figure}

The above description holds well for many structures. However, while steep enough jets approach the self-similar solution, if the initial core is very far from uniform, the core evolution may be dominated by flattening out, and not resemble a BM evolution. 
An example of this can be seen in a hollow jet, in Fig. \ref{fig:hollow_U_theta}. While this jet forms the same lateral structure outside the core and approaches the same self-similar solution, the flattening of the core does not evolve as part of a BM solution. We do not provide a description of the core evolution of hollow jets before the onset of spreading.

\subsubsection{Evolution of a jet with a wide structure}
We now extend our discussion to jets  initial wings that are shallower than $E_{iso}\propto\theta^{-3}$, but still steeper than $E_{iso}\propto\theta^{-2}$, so that the energy integral is convergent. We restrict the discussion to power-law jets with $2<b<3$, although the arguments can be easily applied to jets with a more general wide structure. 

Wide structure jets do not converge to the same self-similar solution reached by narrow structure jets. However, many aspects of the jet evolution remain the same as for narrow jets, and we can generalize our model for narrow structure jets to adapt it to wide structure jets. 
\begin{figure}
    \centering
    \includegraphics[width=\columnwidth]{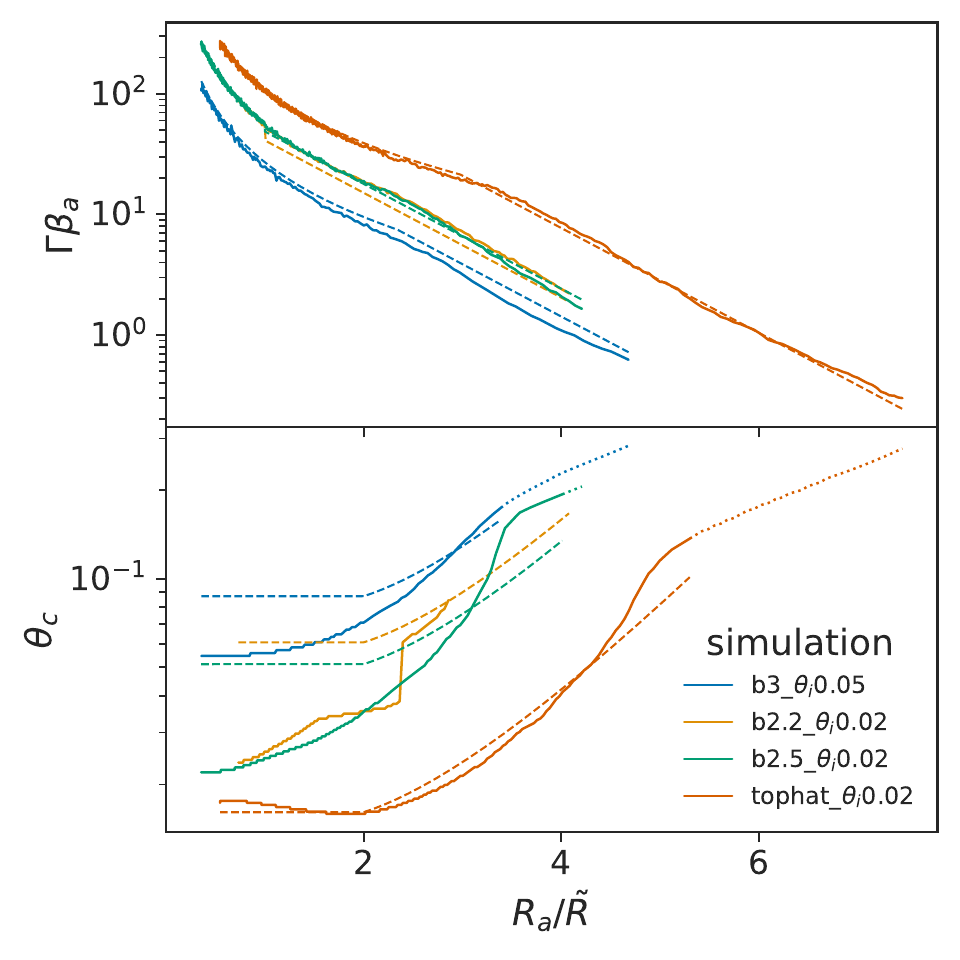}
    \caption{Similar to Fig. \ref{fig:U_a,th_c}, comparing wide jets to narrower structures.}
    \label{fig:wide_Gba_qc}
\end{figure}
As can be seen in Figure \ref{fig:wide_Gba_qc}, like narrow jets,  the proper velocity along the axis initially decreases as part of a BM solution during the pre-spreading phase, where for this phase we use $E_{iso,i,a}$ (and not $\Eiso$) in Eq. \eqref{eq:Gbm}. Similarly, once spreading starts $\Gba$ decreases exponentially while $\theta_c$ grows exponentially. In order to describe this evolution we need to find the expression for $\Rt$ for wide structure jets. Eq. \eqref{eq:Rt} does not give a good approximation, since the energy in the wings is significant and its effect on the evolution of $\Gba$ is different than in the self-similar evolution of narrow jets. Thus,  using the true value of $E_{tot}$ for  wide structure jets in Eq. \eqref{eq:Rt} we obtain value of $\Rt$ that is too large, as if some of the energy in the wings does not contribute to the propagation of the matter along the axis. To adjust further for that, we take the energy in Eq. \eqref{eq:Rt} to be the one that a jet would have if its wings had $b=3$. Formally, since the fraction of the energy in the core of a power-law jet with index $b$ is 
\begin{equation}
    f_c(b) \simeq 1-1 \left(\frac{b}{b-2}\right)^{1-\frac{b}{2}}.
\end{equation}
we replace $E_{tot}$ in equations  \eqref{eq:Rt} and \eqref{eq:qsp} by $E_{tot}\cdot \frac{f_c(b)}{f_c(3)}$.  Figure \ref{fig:wide_Gba_qc} shows that this is a reasonable approximation of $\Gba$ and a less accurate, but still reasonable, approximation of $\qc$. This approximation could be adjusted by re-calibrating the value $\Gbs\qsp$, which we generally do not expect to remain the same. However, we do not find this value changes very significantly, and therefore, for the sake of simplicity, we do not re-calibrate it.

\begin{figure}
    \centering
\includegraphics[width=\columnwidth]{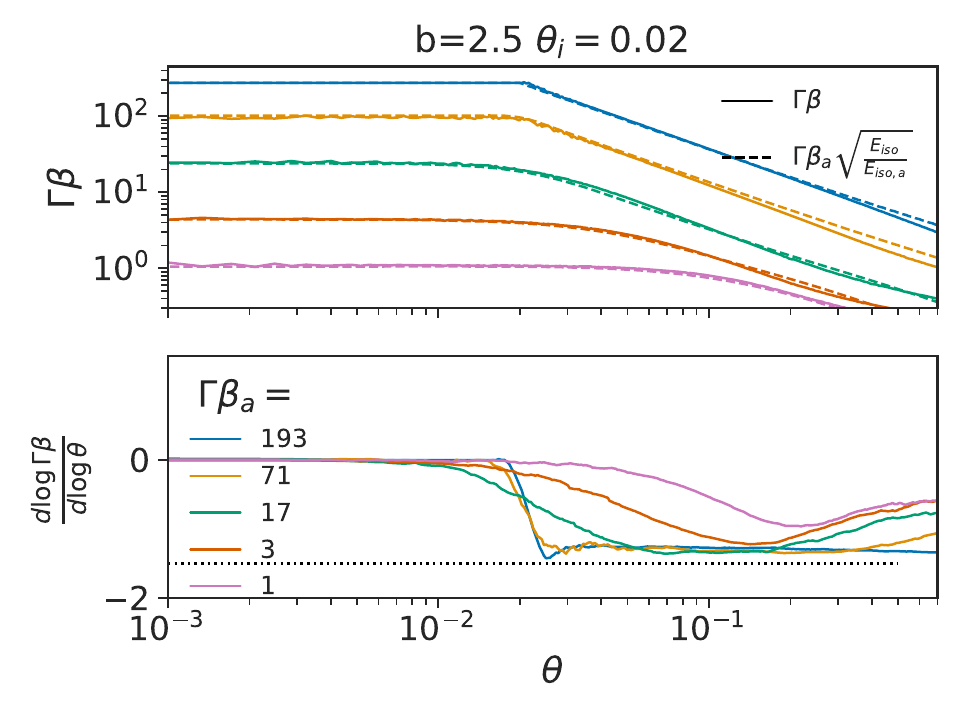}
    \caption{he evolution of $\Gb(\theta)$ and $\frac{d\log\Gb}{d\log\theta}$  of a b=2.5 power-law jet. As opposed to a narrow jet, here the profile outside the core does not evolve significantly, and $E_{iso}\propto\Gamma\beta^2$ at all angles.}
    \label{fig:b2.5}
\end{figure}

The evolution of the structure of wide jets is also different than that of narrow jets. 
Instead of approaching the self-similar solution they retain the initial power-law in the wings throughout the whole evolution, both before and after spreading starts. In addition,  $E_{iso}(\theta)\propto\Gamma^2(\theta)$ at all angles (including the wings) and at all times. The only change to the structure, in this case, is that the transition between the core and the wings smooths out, at the onset of spreading, it reaches approximately: 
\begin{equation}\label{eq: E_wide}  E_{iso}\propto \Gamma^2(\theta) \propto \left(1+\frac{2}{b-2}\left(\frac{\theta}{\theta_c}\right)^2\right)^{-b/2}
\end{equation}
After spreading starts the profile continues to smooth out as it evolves (i.e., there is no single profile that fits the entire evolution). However, taking the power-law outside of the core to be the same as the initial one provides a reasonable approximation.  In Fig. \ref{fig:b2.5} this evolution is shown for a b=2.5 power-law jet. 

The shock radius follows the following relation:
\begin{equation}
    1-\frac{R(\theta)}{ct} \simeq
    \begin{cases}
\frac{1}{8\Gb(\theta)^2} & R_a <\Rs\\
\left(1-\frac{R_a}{ct}\right)\cdot\left(\frac{\Gb}{\Gba}\right)^{-2} & R_a\ge\Rs
\end{cases},
\end{equation}
where the relation between $R_a$ and $t$ can be found by integrating over $\beta_a$. This relation is BM-like at every angle until the onset of spreading and then reaches the relation expected given the relationship between the energy and $\Gb$.


\section{Implications for the emission}\label{sec:emission}
The initial jet structure as well as the structure evolution shape the afterglow emission. Below we discuss the implications of our findings to the emission seen by on-axis and off-axis observers. We also compare the results of our 2D simulations to semi-analytic light curve modeling. 

\subsection{The post jet-break decline rate for on-axis observers}
For jets observed on-axis ($\theta_{obs}<\qco$), the initial core structure shapes the emission until the jet break, and the shape of the jet break is affected by the onset of spreading. The Lorentz factor decline rate in the jet core during the spreading phase together with the spreading of the core, sets the asymptotic decline rate of the light curve after the jet break. \cite{Sari1999} have shown that the combination of an exponential drop in $\Gamma$ and an exponential growth of $\theta_{c}$, such that $\Gamma_a\theta_c \approx 1$, implies an asymptotic post break power-law index $-p$ for observed frequencies $\nu_a<\nu_{obs}$. We find that the core angle spread more slowly than the Lorentz factor decline so $\Gamma_a\theta_c \propto R_a^{-3/2}$ during the spreading phase, and therefore the power-law index of the decline is expected to be steeper than $-p$. This expectation is confirmed by the simulations where we find that for $p=2.2$ the asymptotic decline index is about $-2.4$. This result is consistent with the numerical findings of \cite{vanEerten2013}. We note however that typically for $\theta_{c,0} \gtrsim 0.05$ rad this asymptotic decay is never observed, unless the observer viewing angle is almost along the axis (i.e., $\theta_{obs}\ll \theta_{c,0}$). The reason is that immediately after the jet break there is an "overshoot" and the decay is steeper than the asymptotic one (e.g., for $p=2.2$ immediately after the break we find $d\log(F_\nu)/d\log(T)\approx -2.6$). Now, for observe that is not exactly along the axis the light curve approaches the asymptotic decay index over a duration that is long enough so the transition to the sub-relativistic phase starts affecting the observed emission before the asymptotic decay rate is achieved. Finally, the expectation is that the emission seen by on-axis observers is almost insensitive to the initial structure of the jet. Our results verify this expectation.

\subsection{The rising phase power-law index}
For off-axis observers ($\theta_{obs} \gtrsim 2 \qco$) there is a rising phase, during which the emitting zone scans through the jet wings. Therefore, this phase may depend on the jet initial structure.  If the  emission during the rise is dominated by the self-similar phase of the jet evolution, then we expect the light curve for all initial structures (of narrow structure jets) to coincide. As it turns out, for reasonable jets and observing angles, this is hardly the case and the transition of the jet to the self-similar phase take place only after the peak. To assess when the self-similar solution affects the rise of the light curve, we can calculate for which ratio of $\theta_{obs}/\theta_c$ the peak coincides with the onset of the spreading phase. 
The self-similar phase only starts when $\Gamma_a\theta_c\simeq0.38$, and the light curve peaks when $\Gamma_c(\theta_{obs}-\theta_{c,p})\simeq1.8$, where $\theta_{c,p}$ is the core angle at the lab time that dominates the contribution to the light at the peak of the light curve (see \cite{GovreenSegal2021}). Plugging in  $\Gamma_c\simeq0.5\Gamma_a$, as we find in the self-similar solution, we find that for $\theta_{obs}\lesssim 10\theta_{c,p}$, the light curve peaks before the self-similar evolution starts dominating the emission. Thus, for most observable off-axis jets we expect the rising phase to depend on the jet initial conditions.

\begin{figure}
    \centering
    \includegraphics[width=\columnwidth]{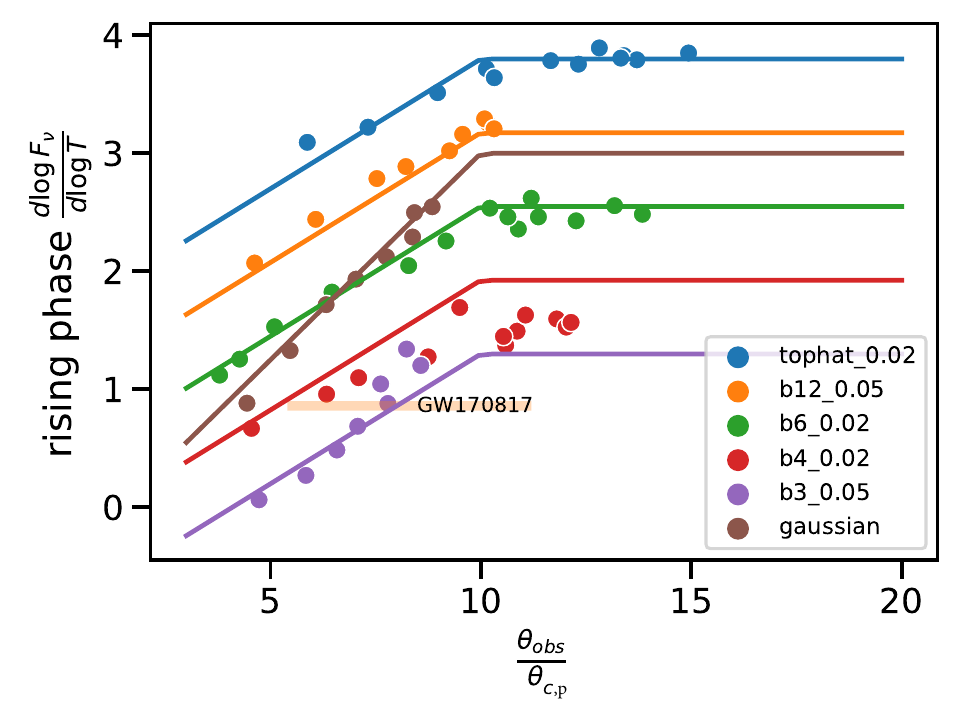}
    \caption{The light curve power-law index during the rising phase as a function of the ratio between the observing angle, $\theta_{obs}$, and the core angle at the time of the peak, $\theta_{c,p}$. The solid lines are the approximation given in Eq. \eqref{eq:rising_pl}, with the calibration constants listed below it. The orange shaded region marks the observational constraints on the rising power-law and angle ratio of GW170817 (see text for details).}
    \label{fig:rising_pl}
\end{figure}

While we do not have an analytical description for the dependence of the light curve rising phase on the evolving jet structure, we can numerically scan the phase space of parameters, and find the best-fit power-law for the rising phase, as a function of the jet initial structure, and the ratio $\theta_{obs}/\theta_{c,p}$. We consider observed frequencies in the range $\nu_a,\nu_m\le \nu_{obs} \le \nu_c$.
Focusing on power-law and Gaussian jets, we find that for every jet structure the rise becomes steeper the farther off-axis the observer is, and reaches a constant value for an observer with $\theta_{obs}>10\theta_{c,p}$. Fitting the rising phase in observed frequency $\nu_a,\nu_m\le \nu_{obs} \le \nu_c$ during the period $T\lesssim 0.5 T_p$ we find that for power-law jets: 

\begin{equation}\label{eq:rising_pl}
    \frac{d\log F_{\nu}}{d\log {T}}(T\le 0.5T_p)=\begin{cases}
        a_1\cdot \left(\frac{\theta_{obs}}{\theta_c}-10\right)+a_2 &\theta_{obs}<10 \theta_{c,p}\\
        a_2 & \theta_{obs}\ge10 \theta_{c,p}.
    \end{cases}
\end{equation}
where $a_1=0.22$ for all power-law jets, and $a_2$ depends on the power-law. We find that $a_2=3.8-7.5/b$ is an excellent approximation. For a Gaussian jet, $a_1=0.35$ and $a_2=2.8$.  The values of $\frac{d\log F_{\nu}}{d\log {T}}(T\le 0.5T_p)$ from this relation are accurate to within $\pm 0.25$. The quality of this approximation can be seen in Fig. \ref{fig:rising_pl}. This figure also shows the parameters estimated for GW170817, which we discuss in \S\ref{sec:170817}.

\subsection{Comparing to afterglowpy}
We compare light curves obtained using our full hydrodynamic simulations to those of afterglowpy \citep{Ryan2020}, as it is the most commonly used semi-analytical scheme. Other commonly used semi-analytic modeling \citep[e.g.,][]{soderberg2006,Lazzati2018} use the non-spreading approximation, which is similar to the non-spreading option in afterglowpy. We calculate the light curve as a post-processing of hydrodynamic simulations using the same method described in \cite{GovreenSegal2023}.

\begin{figure}
    \centering
    \includegraphics[width=\columnwidth]{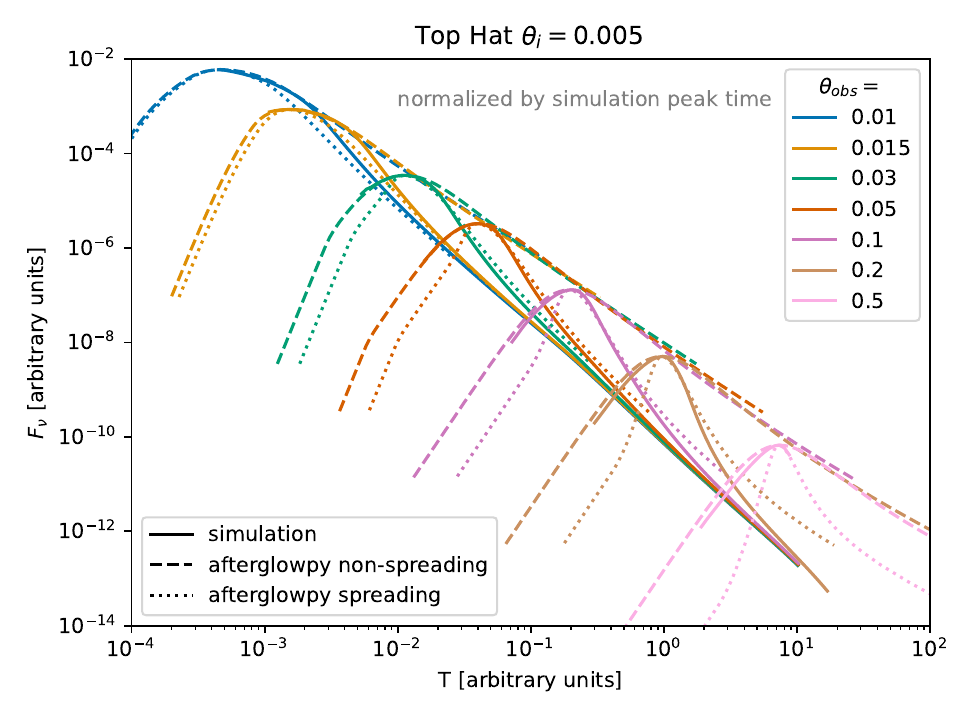}
    \caption{A comparison between light curves calculated using afterglowpy, in spreading and non-spreading modes, and a full 2D simulation. All curves are of  a top hat  jet ($\theta_i=0.005$ rad) observed at various viewing angles, where at all times the observed frequency is in the segment $\nu_a,\nu_m < \nu_{obs}<\nu_c$. The light curves at each viewing angle are all normalized so they coincide at the peak.}
    \label{fig:tophat_afterglowpy}
\end{figure}

\begin{figure}
    \centering
    \includegraphics[width=\columnwidth]{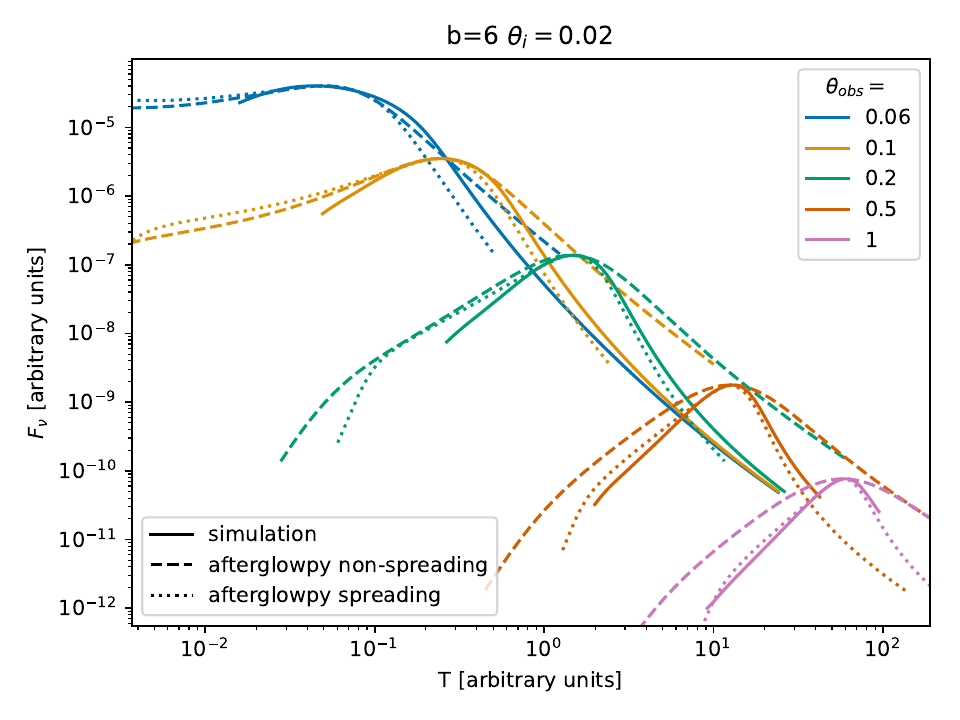}
    \caption{The same as Fig. \ref{fig:tophat_afterglowpy} for a $b=6,\theta_i=0.02$ rad jet.}
    \label{fig:b6_afterglowpy}
\end{figure}

In the following examples, we plot the comparison, for an observed frequency $\nu_m,\nu_a\le \nu_{obs}\le \nu_c$, where $\nu_a,~\nu_m,~\nu_c$ are the self absorption, typical synchrotron and cooling frequencies respectively.
Fig. \ref{fig:tophat_afterglowpy} shows the light curves of a top hat jet with $\theta_i=0.005$ rad observed at various angles, alongside afterglowpy light curves, using both the spreading and the non-spreading options. Afterglowpy light curves are normalized to fit the peak of the light curve based on our simulations.
The shape of the light curve is well described by the non-spreading approximation before the time of the peak, while the spreading approximation is usually good for the decline, though it seems to approach an asymptotic decline that is slightly too shallow and transition to this decline slightly earlier, most prominently at $\theta_{obs}/\theta_c\gtrsim 20$, an angle ratio that is most likely too faint to be observed. Our conclusion is that afterglowpy can provide a very good approximation to top hat jet light curves, if the non-spreading option is used before the peak and the spreading option is used after the peak. 
A similar comparison to a power-law jet with $b=6$ is shown in Fig. \ref{fig:b6_afterglowpy}. The afterglowpy scheme is set up with $\theta_w$ as the angle in the initial conditions at which the matter becomes Newtonian. Here the non-spreading mode is much less accurate while the spreading mode seems to capture the jet evolution reasonably well. The main deviations is that at angles that are more likely to be observed  ($\frac{\theta_{obs}}{\theta_i}\lesssim 20$), afterglowpy predicts rising slopes that are systematically too shallow and peak durations that are systematically is too short. 
We conclude that afteglowpy gives a good approximation to the light curve of structured jets when the spreading option is used, however the differences between afterglowpy light curves and those obtained by the full hydrodynamic model are not negligible and can lead, for example, to an overestimate of the value of $b$ that is needed to fit a given rising light curve. Thus, the errors on the model parameters based on the fit of afterglowpy light curve to the data are most likely to be dominated by systematic errors inherent to the inaccuracy of afterglowpy and not by statistical error.

\section{Implications for the the jet in GW170817}\label{sec:170817}
Our results can be applied to the observations of the afterglow of GW170817 to constrain its jet structure. The geometry of the jet, $\theta_{obs}$ and $\theta_{c,p}$, was measured using the combination of the light-curve and flux centroid motion \citep{Mooley2018,Ghirlanda2019,Mooley2022,GovreenSegal2023}. Figure \ref{fig:rising_pl} shows the constraints on GW170817 in the plane of the rising phase power-law index and the ratio $\theta_{obs}/\theta_{c,p}$. The power-law index, $0.86 \pm 0.04$, is from \cite{Makhathini2021} and the constraint on $\theta_{obs}/\theta_{c,p}$ is from  \cite{GovreenSegal2023} that measure $\theta_{c,p}=1.5-4\deg$ and $\theta_{obs}-\theta_{c,p}=16.8\pm 2 \deg$. This figure shows that the jet must have initially a relatively shallow structure outside of the core (at least at angles that dominated the light during the rising phase). Specifically, if the jet had power-law wings then  $b \simeq 3-4$. A Gaussian jet is also marginally consistent with the observations, although it is less favored. We stress that the jet must have this structure after its emergence from the merger ejecta and before it interacts with the inter-stellar medium, and therefore the origin of the jet structure must be a combination of its structure upon injection from the engine and the jet propagation in the merger ejecta. Interestingly, simulations of the propagation of hydrodynamic and weakly magnetized jets in merger ejecta find that mixing processes during the propagation induce a power-law structure with $b$ in this range \citep{Gottlieb2020,Gottlieb2021}. Jet with steeper initial wings can be ruled out, implying for example that the observations cannot be a result of a top hat jet that developed its structure during its propagation in the interstellar medium (as suggested by \citealt{Ramandeep2019}).   

Our results can be used also to constrain the initial opening angle of the jet. \citealt{GovreenSegal2023} find that at the time of the peak, $\Gamma(\theta_{c,p})(\theta_{obs}-\theta_{c,p})\simeq 1.8$ (see their appendix E1). Thus, the measured value of $\theta_{obs}-\theta_c$ implies that the shock Lorentz factor at the edge of the core at the time of the peak is $\Gamma(\theta_{c,p})\simeq6.2$,  which in turn means that $\Gamma_c\theta_{c,p}=0.16-0.43$. Assuming that the peak took place in the self-similar regime (we check that later) we can use Eqs. \eqref{eq:Gsp} and \eqref{eq:qc}, with the relation $\Gb(\theta_c)=0.46\Gba$, $\Gbs\theta_{\rm c,sp}=0.38$ and $\Rs=3.6 \Rt$, to find that at the time of the peak $R_a \simeq (1-4) \Rt$. The lower limit, $R_a \simeq \Rt$, is obtained for $\theta_{c,p}=4\deg$, and it corresponds to a peak that takes place before the spreading starts. During that time the structure is not self-similar and therefore the measured value of $R_a$ cannot be fully trusted, but in any case the implication is that if $\theta_{c,p}=4\deg$ then we see the peak before spreading starts and thus $\qco\simeq4\deg$ as well. For $R_a\simeq 4\Rt$, which corresponds to $\theta_{c,p}=1.5\deg$, the peak is at the beginning of the self-similar phase and therefore the estimate of $R_a$ can be trusted. From Figure \ref{fig:U_a,th_c_all} we see that for a power-law jet with $b=3$, $\qc(R_a/\Rt=4) \approx 4 \qco$. Thus, if $\theta_{c,p}=1.5\deg$ then $\qco \approx 0.4\deg$. We conclude that the initial opening angle of the jet in GW170817 can be constrained from observations to:
\begin{equation}
    \qco \simeq 0.4-4.5\deg
\end{equation}

\section{Conclusions}\label{sec:Conclusions}
We use analytic arguments alongside a set of 2D relativistic fluid dynamics simulations to investigate the fluid dynamical evolution of relativistic jetted blast waves. Our findings are as follows:
\begin{itemize}
  \item As predicted by many previous studies \citep[e.g.,][]{Rhoads1997,Rhoads1999,Sari1999}, the hydrodynamic relativistic evolution has two phases. During the first, pre-spreading phase, the evolution is as a part of the BM solution. During the second, spreading phase, the shock Lorentz factor drops exponentially with the radius. We find that the transition to the spreading phase takes place roughly when $\Gbs \simeq 0.2 \theta_{c,0}^{-1}$, implying that for jets with initial opening angle $>0.1$ rad the blast wave is mildly relativistic during the transition. 
  \item The core opening angle is roughly constant during the pre-spreading phase and its asymptotic growth during the spreading phase is exponential. During both phases $\Gba \theta_c \propto R_a^{-1.5}$, with a different normalization in each phase. While the asymptotic growth of the core angle is exponential, its expansion at early times of the spreading phase is mitigated by a power-law term ($\qc \propto R_a^{-3/2}\exp [R_a/\Rt]$). As a result the core angle remains much smaller than unity (i.e., the blast wave remains narrowly collimated) also when the shock enters the sub-relativistic regime. To a good approximation, the core opening angle at the time that the shock along the axis becomes mildly relativistic satisfies $\theta_c(\Gba = 1) \simeq 0.4 \theta_{c,0}^{1/3}$.
  \item The profile of jets with narrow initial structure ($E_{iso,i}(\theta)$ in the wings drops faster than $\theta^{-3}$) starts to evolve outside of the core  already during the pre-spreading phase. The wings develop a power-law structure that gradually approaches $\Gb(\theta) \propto \theta^{-1.5}$, and the entire structure evolves to a self-similar profile by the beginning of the spreading phase. The energy profile in the wings during the self-similar phase is $E_{iso} \propto \theta^{-2.6}$.  All narrow structure jets that we studied, regardless of their initial structure (including Gaussian and hollow jets), evolve to the same self-similar profile.
  \item Jets with initial wide structure  ($E_{iso,i}(\theta)$ drops more slowly than $\theta^{-3}$) roughly keep  their initial wing profile throughout the evolution, and they do not evolve to the self-similar profile. Nevertheless, the evolution of $\Gba$ and $\theta_c$ in these jets is largely similar to these of narrow jets.
 \item We show that for off-axis observers, at reasonable observing angles, the self-similar phase contributes to the light curve only after the time of the peak, and therefore the rising of the light curve depends on the initial conditions. We numerically scan parameter space and provide simple formula that relate light curve rising rate to the initial structure and observing angle for observed frequency $\nu_a,\nu_m<\nu_{obs}<\nu_c$.
  \item We compare light curves obtained by the full hydrodynamic simulation to  semi-analytic modeling of afterglowpy, for observed frequency $\nu_a,\nu_m<\nu_{obs}<\nu_c$. We find that it provides good fits when the following modes are chosen. For top hat jets the rising phase should be approximated with the non-spreading mode while the decaying phase (after the peak) should be approximated with the spreading mode. For structured jet it is better to use the spreading mode at all times. We stress, though, that while the approximation of the semi-analytic modeling is reasonable, its inaccuracy is still expected to be the main source of error in fits to well observed cases (such as GW170817). Thus, error estimates that are based on methods such as MCMC, which do not account to systematic errors of the model, cannot be trusted.
   \item We apply our results to learn about the properties of the jet in GW170817 after it emerged from the merger ejecta and before its interaction with the interstellar medium. We find that the rising phase of the light curve implies that the jet initial structure had a relatively shallow wings, such as a  power-law with $b=3-4$. A Gaussian initial profile is marginally consistent with the rising phase. The initial core angle of the jet was $0.4-4.5~\deg$.  
\end{itemize}
 
Based on our results we provide an approximation of the evolution of power-law jets (Appendix \ref{appendix:structure_approx}) which can be used for an improved modeling of the light curve, centroid motion and polarization. In future studies we plan to explore jetted blast waves in other density profiles as well as the transition from the mildly relativistic narrowly collimated blast wave to the spherical Newtonian Taylor–von Neumann–Sedov  blast wave.

\section*{Acknowledgements}
We thank Uri Keshet and Jonathan Granot for useful discussions. This research was partially supported by a consolidator ERC grant 818899 (JetNS) and by an ISF grant (1995/21). TGS thanks the Buchman Foundation for their support.  

\section*{Data availability}
The data underlying this article will be shared on reasonable request to the corresponding author.

\bibliographystyle{mnras}
\bibliography{main}

\appendix
\section{Structure approximation for power-law jets}\label{appendix:structure_approx}
Below we provide formulae that approximate the evolution (global and lateral structure) of jets with an initial power-law profile:
\begin{equation}\label{eq: Eisoi}
    E_{iso}=E_{iso,i,a} 
    \begin{cases}
    1 &\theta<\theta_i\\
    \left(\frac{\theta}{\theta_i}\right)^{-b} &\theta \geq \theta_i\\
    \end{cases}~~,
\end{equation}
where $b$ is constant (the power-law index).
A rough approximation can also be obtained for a top hat jet by using our equations and taking $b$ to infinity. 
We provide all the quantities that are needed to calculate the resulting light curve using the thin shell approximations (using standard afterglow theory; \citealt{Sari_piran_Narayan1998}). 


To describe the structure, our first step is to find the evolution of $\Gba$ and of $\theta_c$, and then use these to describe the structure of $\Gb(\theta),~R(\theta)$ and $M(\theta)$. 
The  total energy and average isotropic equivalent energy of the profile in Eq. \ref{eq: Eisoi} are:
\begin{equation}
    E_{tot}=E_{iso,i,a}\frac{\theta_i^2}{2}\frac{b}{b-2}
\end{equation}
\begin{equation}
    \Eiso=E_{iso,i,a}\frac{b-2}{b-1}
\end{equation}
Using these two expressions, we can now write the typical e-folding radius in the spreading phase: 
\begin{equation}
    \Rt = 0.65 \left(\frac{E_{tot}}{\rho c^2}\right)^{1/3}~,
\end{equation}
the radius at the onset of spreading:
\begin{equation}
    \Rs = 3.6\Rt~,
\end{equation}
the core angle before spreading:
\begin{equation}
    \qco = 0.8\frac{\sqrt{b(b-1)}}{b-2}\theta_i~,
\end{equation}
and at the onset of spreading:
\begin{equation}
    \qsp = 1.6\frac{\sqrt{b(b-1)}}{b-2}\theta_i~.
\end{equation}
Lastly, we need the value of $\Gbs$:
\begin{equation}
    \Gbs=\frac{0.24(b-2)}{\theta_i\sqrt{b(b-1)}}.
\end{equation}

We can now write the expressions for $\Gba$ and $\qc$ as a function of $R_a$:
\begin{equation}\label{eq:Gba_appendix}
\Gamma\beta_{a}=\begin{cases}
\sqrt{\left(\frac{17}{8\pi}\frac{E_{iso,i,a}}{\rho c^{2}R_{a}^{3}}\right)-1} & R_{a}\lesssim R_{sp}\\
\Gamma\beta_{sp}\exp\left(-\frac{R_{a}-R_{sp}}{\tilde{R}}\right) & R_{a}\gtrsim R_{sp}
\end{cases}
\end{equation}
and:
\begin{equation}
\theta_{c}=\begin{cases}
\theta_{c,0} & R_{a}\lesssim 2\Rt\\
0.75\qsp\cdot\left(\frac{R_{a}}{R_{sp}}\right)^{-\frac{3}{2}}\exp\left(\frac{R_{a}-R_{sp}}{\tilde{R}}\right) & R_{a}\gtrsim 2 \Rt
\end{cases}.
\end{equation}
This expression is not identical to the one in the main text, as it includes calibration constants, in order to simplify the writing of the structure. 
In both these expressions, we give an approximate value of $R_a$ at the transition to the spreading regime. The exact value at the transition can be found by numerically solving for $R_a$ for which the values in the two regimes are identical (so the quantities are continuous through the transition). Note that at $R_a\lesssim\Rt$, the expressions have additional intersections, while the transition is always at $R_a>\Rt$.

For many uses, we are interested in the time, in addition to the radius. The time is given by:
\begin{equation}
    t = \frac{R_{a,i}/c}{1-\frac{1}{8\Gamma_{a,i}^2}}+\intop_{R_{a,i}}^{R_{a}}\frac{R'_{a}}{\beta_a c}dR'_{a},
\end{equation}
where $R_{a,i}$, $\Gamma_{a,i}$ are the values of $R_a$ and $\Ga$ in the initial conditions and $\beta_a$ is taken from Eq. \eqref{eq:Gba_appendix}. 

We can now move on to define the structures. We describe jets with an initial power-law structure as a power-law jets at all times, with a gradually changing power-law. For this reason, we define the structure in terms of an effective power-law 
\begin{equation}
    \Gb(\theta) = \Gba\left(1+\left(\frac{\theta}{\theta_c}\right)^{\frac{1}{s}}\right)^{-s\alpha_{eff}},
\end{equation}
where the value of $\alpha_{eff}$ is discussed below, and 
\begin{equation}
    s=
    \begin{cases}
    \frac{\Gbs}{2\Gba} &\Gba>\Gbs\\
    \frac{1}{2} &\Gba \leq \Gbs\\
    \end{cases}
\end{equation}
guarantees a continuous transition from the initial sharp transition from the core to the wings to the much smoother transition during the self-similar phase.
The profile of the shock radius is 
\begin{equation}
    R(\theta) = ct-\left(ct-R_a\right)\cdot \left(\frac{\Gb(\theta)}{\Gba}\right)^{-1.73},
\end{equation}
and the approximate mass per solid angle is:
\begin{equation}
    \frac{dM}{d\Omega}=\rho\frac{R(\theta)^3}{3}.
\end{equation}
Note that the mass in fact flows sideways, and does not exactly follow this expression, however, this adds a minor correction to the emission.

The value of $\alpha_{eff}$, depends on the initial structure, and whether there is relativistic matter at angles larger than $\theta_{NR}$.
We approximate $\theta_{NR}$ as constant throughout the evolution:
\begin{equation}
    \theta_{NR} = 0.31\qsp^{1/3}.
\end{equation}
If initially the jet structure is such that $\theta(\Gb=1)\ge \theta_{NR}$, than as long as this requirement hold true:
\begin{equation}
\alpha_{eff}=\frac{b}{2}.
\end{equation}
If $\theta(\Gb=1)\le \theta_{NR}$, either initially or after a while, and as long as $\Gba\ge \Gbs$, $\alpha_{eff}$ is determined by the requirement that $\theta(\Gb=1)=\theta_{NR}$:
\begin{equation}\label{eq:alpha_eff2}
\alpha_{eff}=\frac{\log\left(\Gamma\beta_{a}\right)}{s\cdot\log\left(1+\left(\frac{\theta_{NR}}{\theta_{c}}\right)^\frac{1}{s}\right)}~~~.
\end{equation}
Eqs. \eqref{eq:alpha_eff2} is used up to the point that   $\alpha_{eff}=1.5$, which takes place roughly at $\Rs$. From this point on we use the self-similar structure where $\alpha_{eff}=\alpha=1.5$.

\begin{figure*}
    \centering
    \includegraphics[width=\textwidth]{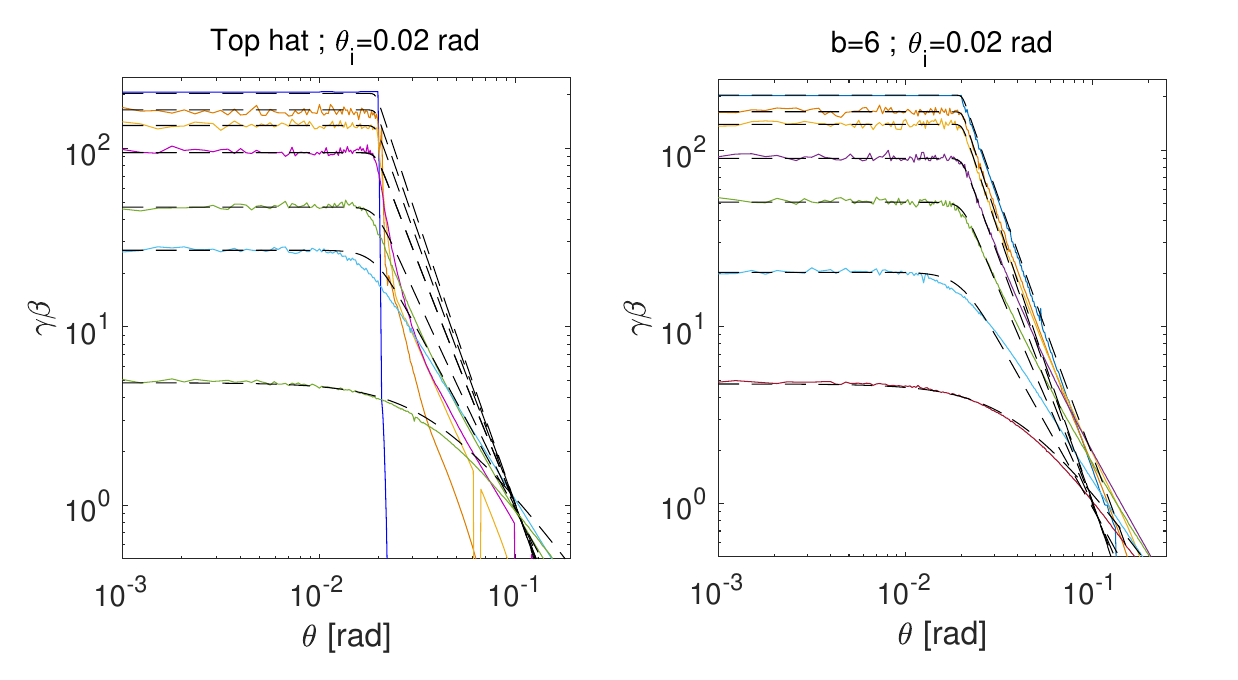}
    \caption{The lateral structure of the freshly shocked material proper velocity, $\gamma\beta(\theta)$, at various stages of the shock propagation. {\it colored solid lines:} results of the full 2D hydro simulations. {\it Dashed black lines:} Semi-analytic formulae, Eqs. \eqref{eq: Eisoi}-\eqref{eq:alpha_eff2}. The {\it left panel} shows a top hat jet and the {\it right panel} shows a $b=6$ power-law jet.   }
    \label{fig:structure}
\end{figure*}

\begin{figure}
    \centering    \includegraphics[width=\columnwidth]{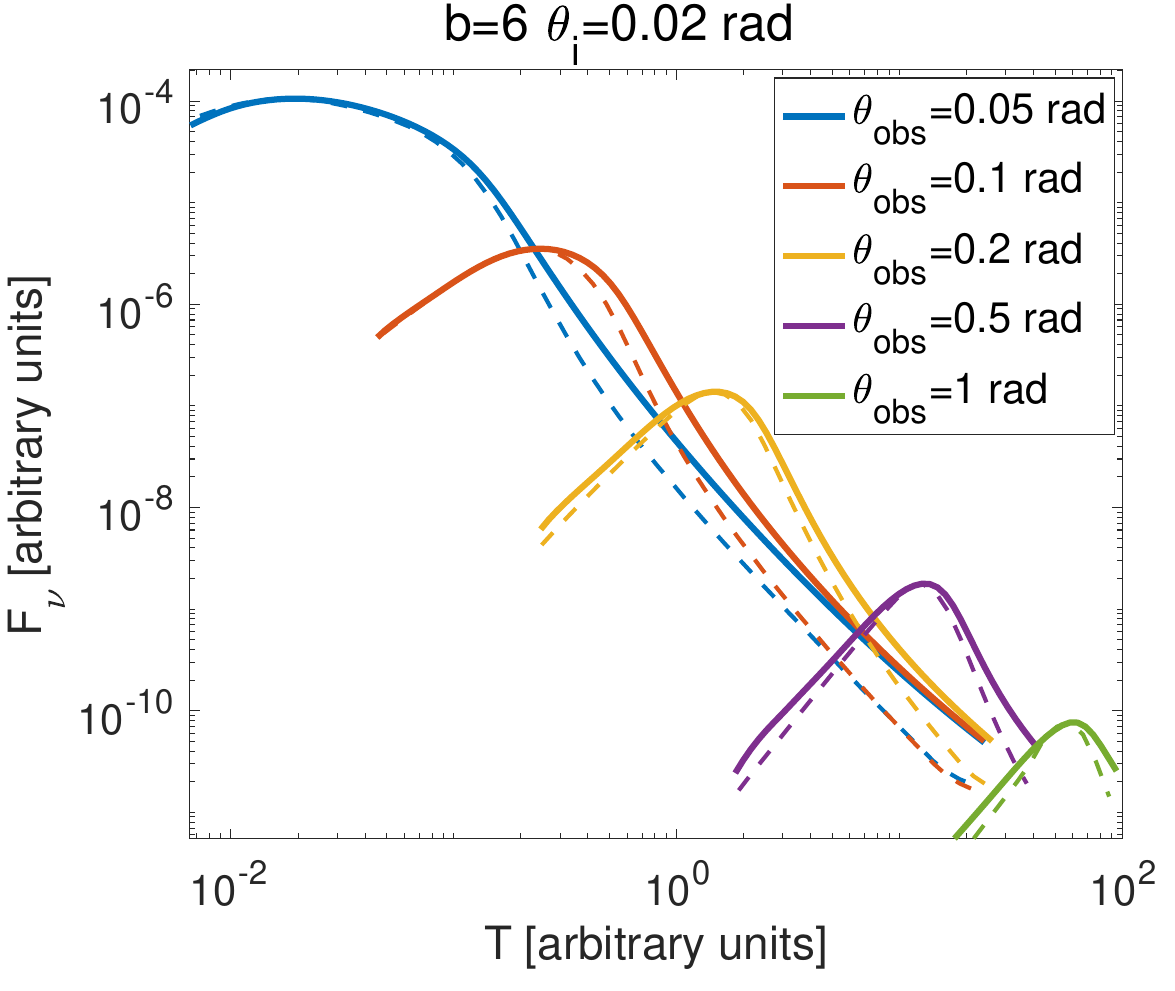}
    \caption{Light curves from the full 2D hydrodynamic simulations of a $b=6$ power-law jet (solid lines) are compared to light curves calculated with the approximated structure (Appendix \ref{appendix:structure_approx}; see text) and the thin shell approximation (dashed lines). The light curves are normalized to coincide at the peak.}
    \label{fig:tophat_LC}
\end{figure}

\begin{figure}
    \centering
    \includegraphics[width=\columnwidth]{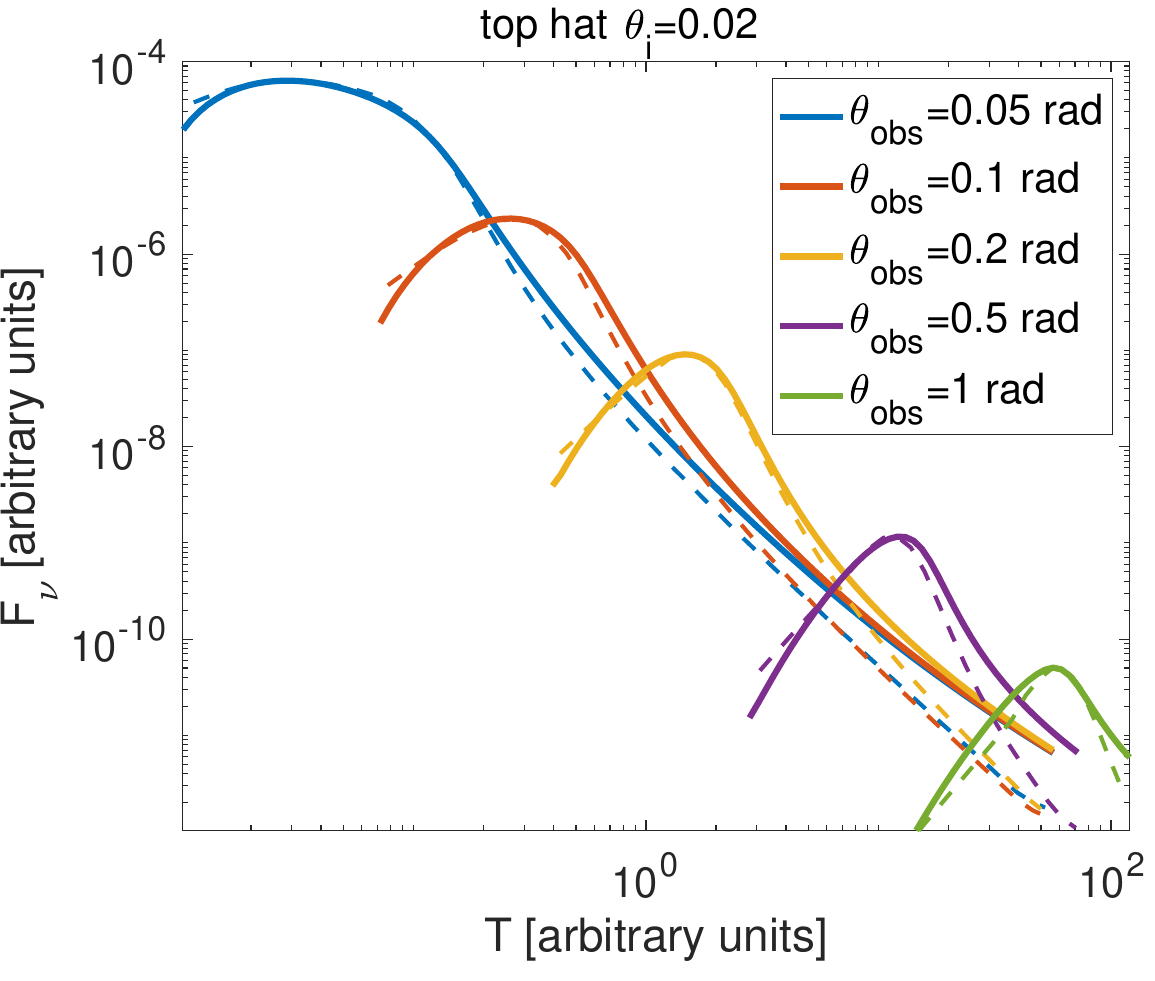}
    \caption{Same as Fig. \ref{fig:tophat_LC}, for a top hat jet.}
    \label{fig:b6_LC}
\end{figure}
Fig. \ref{fig:structure} depicts the structures obtained by Eqs. \eqref{eq: Eisoi}-\eqref{eq:alpha_eff2} during various stages of the evolution of a power law jet with $b=6$ and a top hat jet. It shows also the structure  of the same jets at the same stages of the evolution as obtained with full 2D simulations. It shows that the structure approximation to the $b=6$ power-law jet is very good. The approximation of the top hat jet structure is not as good, but it is still reasonable. The accuracy increases with time as the jet approaches the self-similar profile. 
The quality of the light curves resulting from the structure approximations of the top hat  and a b=6 jets is shown in Figs. \ref{fig:tophat_LC} and \ref{fig:b6_LC}. These figures compare the light curve obtained using an approximated structures to light curves that are based on full 2D simulations.
The figures show that the approximation is good. For $b=6$ it captures accurately the rising phase, much better than the approximation of afterglowpy. Also the shape of the peak is fairly accurate, where the main deviation is an episode with a sharper decay than in the simulation. For a top hat jet the quality of the approximation is roughly similar to that of the $b=6$ jet, where the main difference is a somewhat less accurate approximation of the rising phase.

To summarize, our approximation provide a useful tool for deriving fast semi-analytical scheme that provides more accurate light curves, images and polarization calculations then currently available. Yet, at least for light curves, very accurate results still require the use of full hydro simulations.

\section{Simulation setup}\label{appendix:sim_setup}
Part of the simulations we used here were already presented in  \cite{GovreenSegal2023}, where the setup is described in detail. The setup of the rest of the simulations is almost identical to the one in \cite{GovreenSegal2023} apart for the following modifications. First the range of the AMR parameter is 0.5-2 (compared to 1-2 in \citealt{GovreenSegal2023}). Second, in some simulations, the grid does not extend all the way to $\frac{\pi}{2}$, thus the grid spacing in the $\theta$ direction is defined as follows: 
\begin{equation}
    \theta_{j-\frac{1}{2}}=\theta_{max}\left(f_{R}\frac{j}{N_{\theta}}+\left(1-f_{R}\right)\left(\frac{j}{N_{\theta}}\right)^{5}\right),
\end{equation}
where $N_\theta$ is the number of grid cells in the $\theta$ direction, $f_R$ defines the fraction of the grid in which the cells are smaller, and $\theta_{max}$ is the maximal value of $\theta$ included in the simulation. The values of these parameters are listed in table \ref{tab: sim_setup}. In all simulations with $\theta_{max} < \frac{\pi }{2}$, $\theta_{max}$ is set so that no relativistic material reaches the grid boundary. In simulations with $\theta_{max}=\frac{\pi}{2}$, the boundary condition is changed to reflective before relativistic matter reaches the outer boundary. 
\begin{table}
    \centering
\begin{tabular}{|c|c|c|c|c|c|}
\hline 
\multicolumn{2}{|c|}{Simulation} & $\theta_{i}$ & $N_{\theta}$ & $f_{R}$ & $\theta_{max}$\tabularnewline
\hline 
\hline 
\multicolumn{1}{|c}{\multirow{7}{*}{top hat}} & \multirow{7}{*}{} & $0.005$ & 288 & 0.25 & 0.1\tabularnewline
\cline{3-6} \cline{4-6} \cline{5-6} \cline{6-6} 
 &  & $0.02$ & 400 & 0.07 & $\frac{\pi}{2}$\tabularnewline
\cline{3-6} \cline{4-6} \cline{5-6} \cline{6-6} 
 &  & $0.05$ & 200 & 0.15 & $\frac{\pi}{2}$\tabularnewline
\cline{3-6} \cline{4-6} \cline{5-6} \cline{6-6} 
 &  & $0.1$ & 300 & 0.3 & $\frac{\pi}{2}$\tabularnewline
\cline{3-6} \cline{4-6} \cline{5-6} \cline{6-6} 
 &  & $0.15$ & 300 & 0.225 & $\frac{\pi}{2}$\tabularnewline
\cline{3-6} \cline{4-6} \cline{5-6} \cline{6-6} 
 &  & $0.2$ & 192 & 0.3 & $\frac{\pi}{2}$\tabularnewline
\cline{3-6} \cline{4-6} \cline{5-6} \cline{6-6} 
 &  & $0.3$ & 192 & 0.35 & $\frac{\pi}{2}$\tabularnewline
\hline 
\multirow{9}{*}{Power-Law} & $b=2.2$ & $0.02$ & 240 & 0.07 & $\frac{\pi}{2}$\tabularnewline
\cline{2-6} \cline{3-6} \cline{4-6} \cline{5-6} \cline{6-6} 
 & $b=2.5$ & $0.02$ & 288 & 0.07 & $\frac{\pi}{2}$\tabularnewline
\cline{2-6} \cline{3-6} \cline{4-6} \cline{5-6} \cline{6-6} 
 & \multirow{2}{*}{$b=3$} & $0.05$ & 500 & 0.4 & $\frac{\pi}{2}$\tabularnewline
\cline{3-6} \cline{4-6} \cline{5-6} \cline{6-6} 
 &  & $0.15$ & 600 & 0.6 & $\frac{\pi}{2}$\tabularnewline
\cline{2-6} \cline{3-6} \cline{4-6} \cline{5-6} \cline{6-6} 
 & \multirow{3}{*}{$b=6$} & $0.02$ & 384 & 0.07 & $\frac{\pi}{2}$\tabularnewline
\cline{3-6} \cline{4-6} \cline{5-6} \cline{6-6} 
 &  & $0.05$ & 200 & 0.15 & $\frac{\pi}{2}$\tabularnewline
\cline{3-6} \cline{4-6} \cline{5-6} \cline{6-6} 
 &  & $0.15$ & 500 & 0.5 & $\frac{\pi}{2}$\tabularnewline
\cline{2-6} \cline{3-6} \cline{4-6} \cline{5-6} \cline{6-6} 
 & \multirow{2}{*}{$b=12$} & $0.05$ & 200 & 0.15 & $\frac{\pi}{2}$\tabularnewline
\cline{3-6} \cline{4-6} \cline{5-6} \cline{6-6} 
 &  & $0.15$ & 300 & 0.225 & $\frac{\pi}{2}$\tabularnewline
\hline 
$b6c1$ & $b=1,b=6$ & $0.02$ & 288 & 0.07 & $\frac{\pi}{2}$\tabularnewline
\hline 
\multicolumn{1}{|c}{\multirow{2}{*}{Gaussian}} & \multicolumn{1}{c|}{} & $0.02$ & 288 & 0.07 & $\frac{\pi}{2}$\tabularnewline
\cline{3-6} \cline{4-6} \cline{5-6} \cline{6-6} 
 &  & $0.07$ & 200 & 0.15 & $\frac{\pi}{2}$\tabularnewline
\hline 
\multirow{2}{*}{Hollow} & \multicolumn{1}{c|}{} & $0.05$ & 200 & 0.15 & $\frac{\pi}{2}$\tabularnewline
\cline{2-6} \cline{3-6} \cline{4-6} \cline{5-6} \cline{6-6} 
 & $b=6$ & $0.15$ & 400 & 0.4 & $\frac{\pi}{2}$\tabularnewline
\hline 
\end{tabular}
    \caption{Grid setup parameters for simulations used in this work.}
\end{table}\label{tab: sim_setup}

\section{Adiabatic index effect on the mildly relativistic phase}\label{appendix: EOS}
\begin{figure}
    \centering
    \includegraphics[width = \columnwidth]{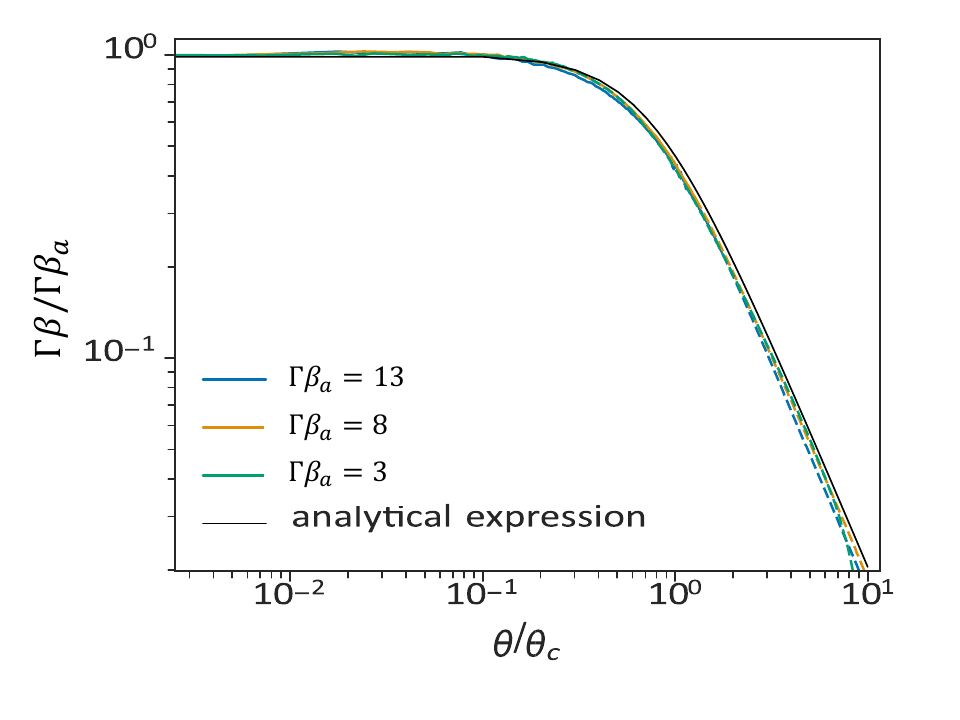}
    \caption{The self-similar solution plotted for a top hat jet with $\theta_i=0.02$ and a Synge-like equation of state. The analytical expression plotted is \ref{eq:U_self-simiar}.  Note that the definition of $\theta_c$ is not applicable to the curve with $\Gba=3$ (our definition of $\theta_c$ does not behave well when $\Gamma_c$ becomes mildly relativistic, see discussion below Eq. \ref{eq:qc_def}). Therefore, the value of $\theta_c$ for this curve is found by the angel at which $\frac{d\log(\Gb)}{d\log(\theta)}=-1$. }
    \label{fig:synge_self-similar}
\end{figure}

The simulations used in the main text all have an adiabatic index of $4/3$, as our focus is the ultra-relativistic phase. However, as we noticed that many results can be naturally extended to the mildly relativistic phase, we consider here the effect of the choice of the adiabatic index, by comparing the simulations to simulations with an adiabatic index that changes smoothly from $4/3$ to $5/3$, in a manner that approximates the Synge gas law (see details on the implementation in \citealt{Ayache2022}). We find that the effect of the equation of state is minor. This can be seen in Fig. \ref{fig:Ua,th_c,synge}, in which $\Gba$ and $\qc$ are plotted as a function of the normalized radius for top hat jets with $\theta_i=0.02$ rad and $\theta_i=0.2$ rad. In the narrower jet there is no noticeable difference while in case of $\theta_i=0.2$ rad there is a small difference during the spreading phase (about $10\%$). When considering the self-similar solution, we find that $\Gb(\theta)$ still follows the same behavior, also when changing the EOS. This can be seen in Fig. \ref{fig:synge_self-similar}, in which the self-similar solution is shown for a top hat jet simulation with $\theta_i=0.02$ and a Synge equation of state. 
\begin{figure}
    \centering
\includegraphics[width = \columnwidth]{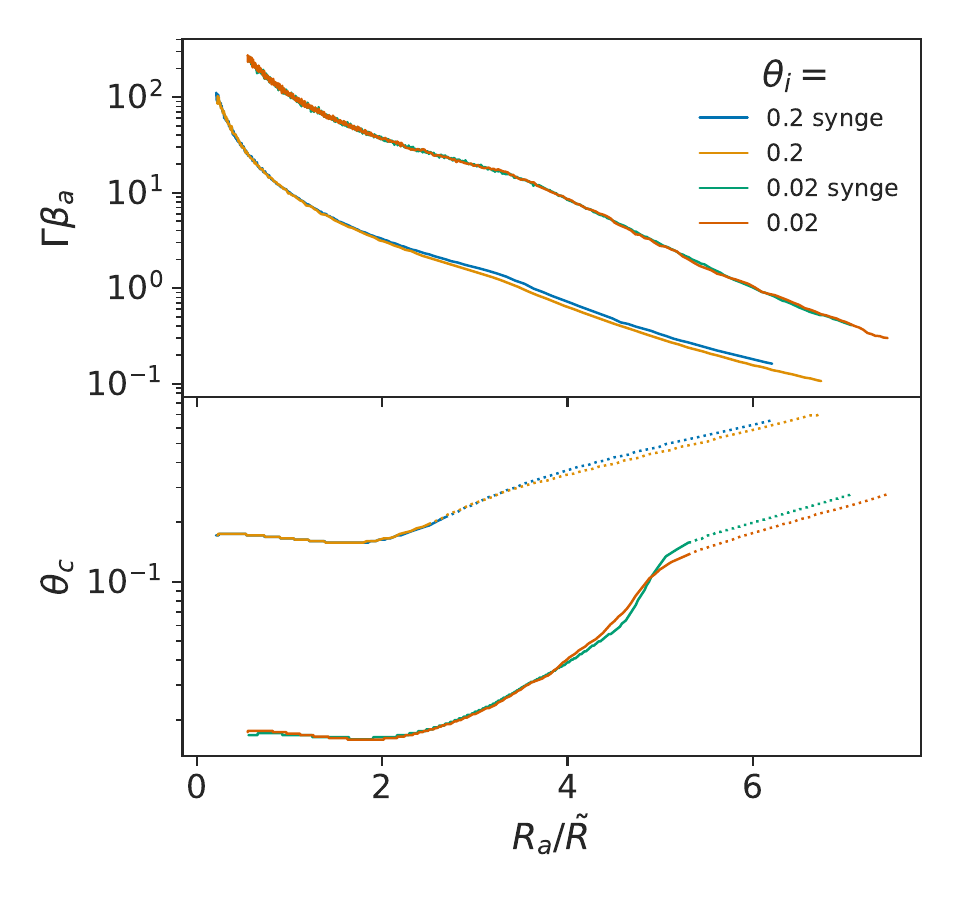}
    \caption{The effect of the equation of state on the evolution of $\Gba$ and of $\qc$ is studied, by comparing simulations of top hat jets with $\theta_i=0.02$ and $0.2$ with an adiabatic index of $4/3$ and a Synge-like equation of state.}
    \label{fig:Ua,th_c,synge}
\end{figure}

\section{A comparison between different definitions of $\theta_c$}\label{appendix: theta_c}
\begin{figure}
    \centering
    \includegraphics[width=\columnwidth]{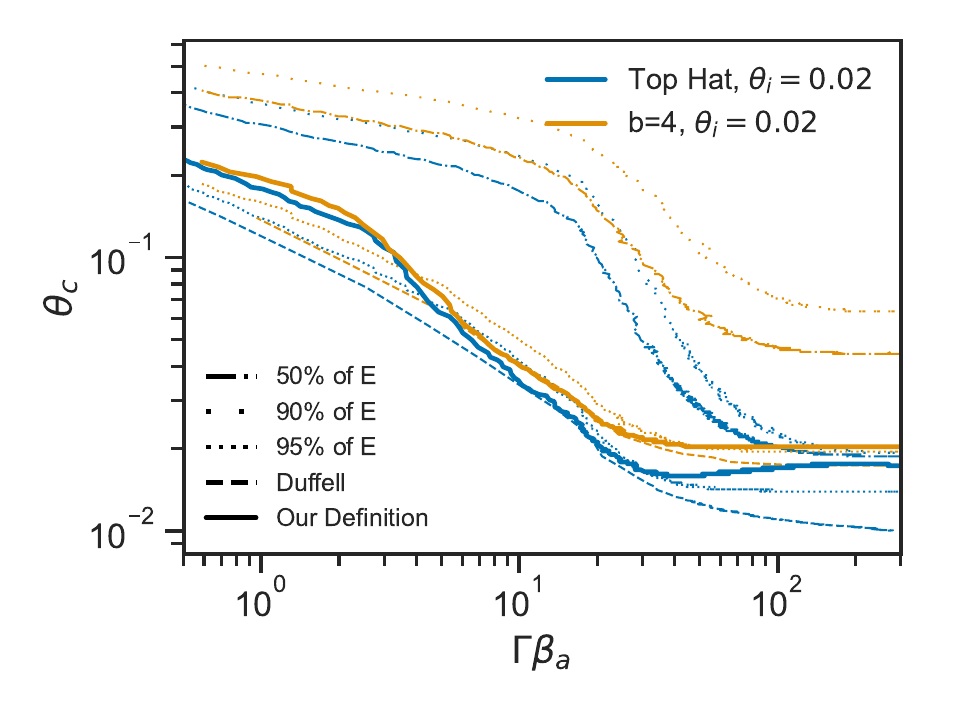}
    \caption{A comparison of different definitions of $\theta_c$. The figure depicts the evolution of $\theta_c$, according to different definitions, with the proper velocity (note that the time flows from large to small values of $\Gba$). It shows that definitions in which $\theta_c$ contains a very large fraction of blast wave energy ($90\%$ and $95\%$) are sensitive to the wings and  are insensitive to the core (see main text for discussion) and therefore, these are not good definitions (at least for the purposes of this paper). All other popular definitions: the region that contains $50\%$ of the energy, the average angle  defined by \citet{Duffell2018} and our definition, are rather similar and can probably serve well for the type of analysis we carry out in this paper.}
    \label{fig:theta_core_comp}
\end{figure}

Figure \ref{fig:theta_core_comp} shows the evolution of $\theta_c$  using various definitions. It shows that the two popular definitions in which the core angle is defined as the region that contains $90\%$ or $95\%$ of the energy are highly sensitive to the shape of the wings and are only weakly sensitive to the properties of the core. As a result, $\theta_c$ according to these definitions grows very rapidly at early times, long before the spreading phase starts, when the wings outside of the core evolve. In contrast, during the spreading phase, $\theta_c$ according to these definitions barley grows, even though the central region where the energy profile is relatively flat (which is defined as the core using the other definitions) grows exponentially. The reason is that during the self-similar phase the wings remains almost unchanged. We conclude that these definitions are too sensitive to the wings and are not capturing the essence of the core evolution well.  
All other definitions, including the region that contains $50\%$ of the energy, the average angle  defined by \cite{Duffell2018} and our definition, are pretty similar and it's likely that for most jets, the analysis used here could be adapted to these definitions. These definitions will differ from ours in several cases. For example, for a hollow jet, in which the core energy is smaller than a top hat jet with the same energy, but the core size relevant for observations is still the same and is correctly recovered using our definition. 

\section{Indications of geometrical self-similarity}\label{appendix: GSS}
In addition to the time-wise lateral self-similarity, we find an interesting property of the jet radial structure during the self-similar phase. This is similar to what \cite{Keshet2015} describe as a geometrical self-similarity - at any given time and at any given angle the radial structure seems to have a self-similar profile.  Below we illustrate these results, which may be useful for finding the self-similar solution analytically (if, as we expect, there is such a solution). 

\begin{figure*}
    \centering
    \includegraphics[width = \textwidth]{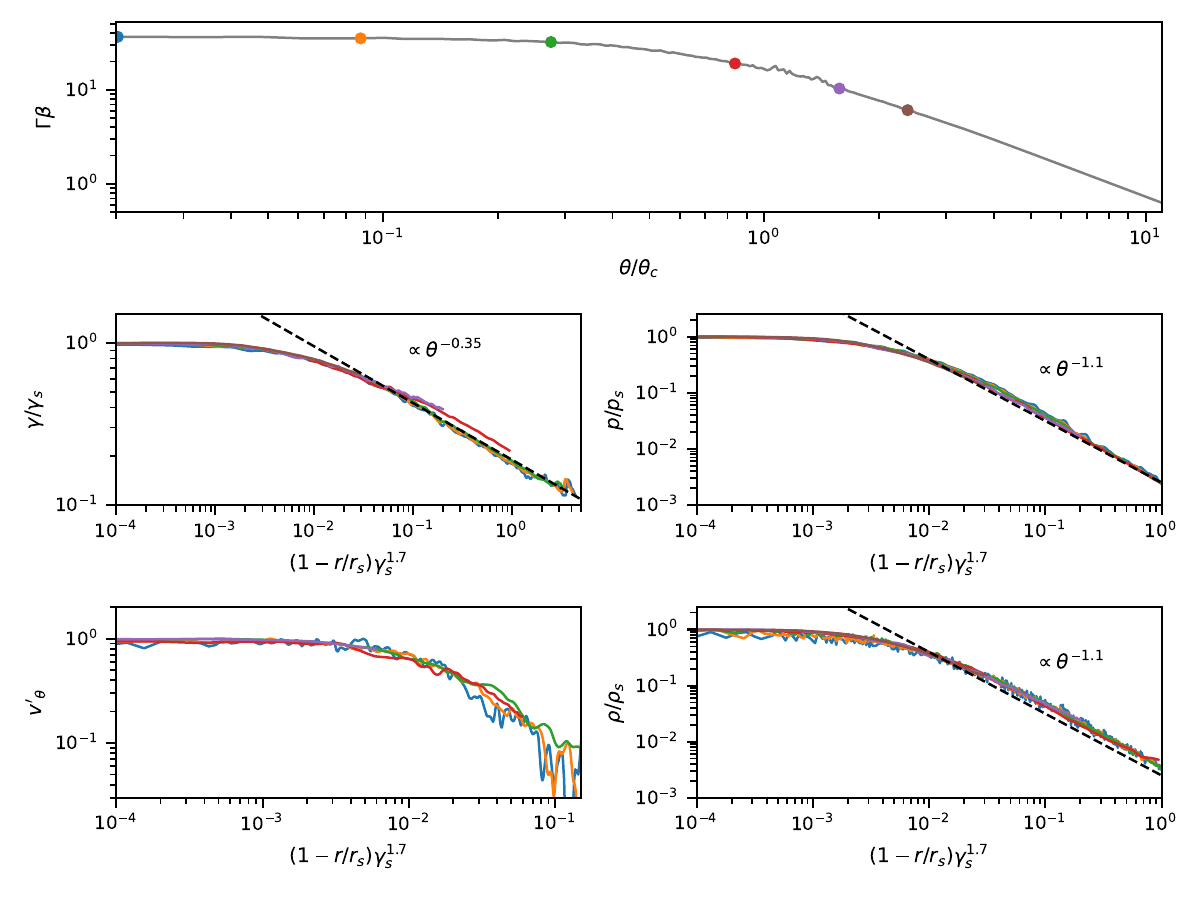}
    \caption{This figure shows indications for geometrical self-similarity, in a snapshot from the simulations, taken during the spreading phase. In the top panel, the shock lateral structure - $\Gb(\theta/\theta_c)$ is plotted, at a snapshot in time , and on it are marked points at which the radial structure is probed. These points are color-coded, and we use the same colors in the following plots. The following plots all show values of fluid dynamical properties, normalized by their value at the shock front as a function of the normalized distance behind the shock. The black dashed lined mark power-laws fitted by eye, and their index is denoted on the graph.}
    \label{fig:gss}
\end{figure*}

We find that during the self-similar phase, at any given time and at any given angle $\theta$, various fluid properties have a similar radial profile when they are scaled by  $(1-\frac{r}{r_s})\gamma_s^{1.7}$, where $\gamma_s$ is the Lorentz factor of the freshly shocked matter and $r_s$ is the shock radius, both taken at the same angle $\theta$. This is illustrated in Figure \ref{fig:gss}. The top panel of this figure shows $\gamma_s(\theta)$ at a given time, and on it in colorful dots are marked various angles, the dot color serves as a legend for the following plots. In the following panel, for each of these angles we select a radial cut, and plot the matter Lorentz factor normalized by the post-shock Lorentz factor as a function of $(1-\frac{r}{r_s})\gamma_s^{1.7}$ for that angle. One may see 
that all the profiles at all angles have a similar structure, specifically, for $r_s\left(1\, -\frac{0.01}{\gamma ^{1.7}}\right) \lesssim r\le r_s$, the Lorentz factor is 
nearly constant and decreases as a power-law in $(1-\frac{r}{r_s})\gamma_s^{1.7}$ for smaller values of $r$. The following panels portray the pressure $p$, the density $\rho$ and the velocity  in the $\theta$ direction, as measured in a frame co-moving with the fluid in the radial direction, $v'_\theta$. $\rho$ and $p$ show a similar behaviour to that seen in $\gamma$, while $v'_\theta$ shows a similar plateau, but can only be plotted over a much smaller range before it becomes dominated by numerical noise, and we cannot conclude it's behavior at larger angles. 

\end{document}